\documentclass[12pt]{article}
\usepackage{cite}
\usepackage{bm}

\usepackage{relsize}

\usepackage[bookmarks, breaklinks, colorlinks,urlcolor=black, citecolor=red, 
linkcolor=blue]{hyperref}

\usepackage{url}

\usepackage{ulem}

\usepackage{graphicx,floatflt,amssymb,rotate,epsfig}
\usepackage[utf8]{inputenc}
\usepackage[inline]{enumitem}
\usepackage{mathrsfs}
\usepackage{amssymb}
\usepackage{amsmath}
\usepackage{color}
\usepackage{graphicx}
\usepackage{subfig}
\usepackage{multirow}
\usepackage{mathtools}
\usepackage{float}
\usepackage{pstricks}
\usepackage[toc,page]{appendix}
\usepackage{pifont}
\usepackage{braket}
\usepackage{bbold}

\usepackage{slashed}

\def\bea{\begin{eqnarray}}
	\def\eea{\end{eqnarray}} 
\def\be{\begin{equation}}
	\def\ee{\end{equation}}

\def\gev{\ensuremath{\mathrm{Ge\kern -0.1em V}}}
\def\mev{\ensuremath{\mathrm{Me\kern -0.1em V}}}

\def\roughly#1{\mathrel{\raise.3ex\hbox
		{$#1$\kern-.75em\lower1ex\hbox{$\sim$}}}}
\def\lsim{\roughly<}

\definecolor{Darkgreen}{RGB}{30,150,30}

\begin{document}

\begin{flushright}
	%SI-HEP-2021-19\\[1cm]
	%\today
\end{flushright}

%%%%%%%%%%%%%%%%%%%%%%%%%%%%%%%%%%%%%%%%%%%%%%%%%%%%%%%

\begin{center}
	
	{\Large\bf 
		Constraining inverse moment of $B$-meson  distribution \\[5pt] amplitude using Lattice QCD data}
	\\[8mm]
	{
		Rusa Mandal$^a$\,\footnote{rusa.mandal@iitgn.ac.in}, Soumitra Nandi$^b$\,\footnote{soumitra.nandi@iitg.ac.in} and Ipsita Ray$^a$\,\footnote{ipsitaray02@gmail.com} 
	}
	\\[10pt]
	$^a$\,{\small\it  Indian Institute of Technology Gandhinagar, Department of Physics,  Gujarat 382355, India} \\[2mm]
	$^b$\,{\small\it Indian Institute of Technology, North Guwahati, Guwahati 781039, Assam, India}

\end{center}
%%%%%%%%%%%%%%%%%%%%%%%%%%%%%%%%%%%%%%%%%%%%%%%%%%%%%%%

\vspace{4mm}

\begin{abstract}
We constrain the inverse moment of the $B$-meson light-cone distribution amplitude (LCDA), $\lambda_B$ in heavy quark effective theory, using form factor estimates from Lattice QCD collaboration. The estimation of the parameter $\lambda_B$ has, until now, relied solely on QCD sum rule methods and deals with significant uncertainty. In this work, we express the form factors for the $B \to K$ channel, calculated within the light-cone sum rule (LCSR) approach, in terms of the $B$-meson LCDAs. By incorporating recent Lattice results from the HPQCD collaboration for the $B \to K$ form factors at zero momentum transfer ($q^2$ = 0), we impose constraints on this parameter. Consequently, we achieve a twofold reduction in uncertainty compared to the QCD sum rule estimate, yielding $\lambda_B=338\pm 68$\,MeV, when the $B$-meson LCDAs are expressed in the Exponential model. Additionally, we compare the form factor predictions, using the constrained $\lambda_B$ value, with the earlier analyses for other channels as well, such as $B\to \pi$ and $B \to D$. 
\end{abstract}



\section{Introduction}
\label{introduction}

	Significant progress has been made in recent years regarding the determination of the hadronic matrix elements of the form $\langle h'|\bar{q}' \Gamma q|h\rangle$, which describe transitions between initial state hadrons $(h)$ and final state hadrons $(h^\prime)$.
	These matrix elements are parameterized in terms of form factors which are non-perturbative objects and serve as dominant sources of uncertainties for the theoretical predictions of observables in the flavor sector. The estimation of form factors relies on different approaches, each demonstrating varying degrees of reliability across different kinematic regions. 
	Lattice QCD calculations provide the most precise estimates of the form factors that are presently most effective in the high momentum transfer i.e., $q^2$ regime. However, the extrapolation to $q^2$ = 0 regime might increase the uncertainties for the data points due to additional assumption entering in the kinematic extrapolation stemming from unitarity and analyticity (see e.g., in Refs. \cite{Boyd:1994tt,DiCarlo:2021dzg,Flynn:2023qmi}). On the other hand, LCSR, QCD factorization/soft collinear effective theory results are used for the low $q^2$ region. LCSR computations employing $B$-meson LCDAs depend on a parameter $\lambda_B$, which is the first inverse moment of the $B$-meson distribution amplitude and is highly uncertain. Currently, the precise estimate of this parameter is not achievable with QCD sum rule technique~\cite{Braun:2003wx,Khodjamirian:2020hob} which requires knowledge of the non-local condensates. The most accurate estimates of $\lambda_B$ may be possible from the measurements of the photoleptonic decay $B \to \gamma \ell \nu$, which is, however, experimentally quite challenging. 
	In an analysis by the Belle collaboration \cite{Belle:2018jqd}, it is found that, the measurement of the branching fraction of $B \to \gamma \ell \nu$ decay is  limited by low signal yield, and a lower limit on $\lambda_B$ $>$ 0.24 $\text{GeV}$ at 90$\%$ confidence interval (CI) is obtained. Discussions from the Lattice QCD perspectives on these radiative meson decays can be found in Ref.~\cite{Giusti:2023pot,Desiderio:2020oej,Frezzotti:2023ygt}. Further attempt has been made on the indirect extraction of $\lambda_B$. For example, in~\cite{Wang:2015vgv}, the LCSR prediction with $B$-meson LCDA of $B\to \pi$ form factor (at zero momentum transfer) is matched with the QCD sum rule prediction of the same form factor computed using pion distribution amplitude. A comparison between LCSR and SCET calculations of $B \to \gamma$ form factors~\cite{Janowski:2021yvz} also provides indirect estimates for $\lambda_B$. Furthermore, a SCET based formulation of the LCSR for $B \to \rho$ form factors is compared to the estimates using NLO LCSR with $\rho$-meson distribution amplitudes in \cite{Gao:2019lta}.
	In Table~\ref{tab:values} we summarize the findings for these different approaches. In the indirect extraction cases, different models of the $B$-meson LCDAs provide different estimates highlighting the model dependence of the extracted parameter $\lambda_B$.
	\begin{table}
		\centering
		\renewcommand*{\arraystretch}{1.3}
		\begin{tabular}{|c|c|c|}
			\hline
			Approach & $\lambda_B$ (MeV) & Details    \\
			\hline\hline
			\multirow{2}{*}{QCD sum rule} & $460\pm 110$~\cite{Braun:2003wx} &  \\
			& $383\pm 153$~\cite{Khodjamirian:2020hob} &  \\ \hline
			\multirow{4}{*}{$B \to \pi$ form factor} & $354^{+38}_{-30}$~\cite{Wang:2015vgv} &  Model-I\\
			& $368^{+42}_{-32}$~\cite{Wang:2015vgv} &  Model-II\\
			& $389^{+35}_{-28}$~\cite{Wang:2015vgv} &  Model-III\\
			& $303^{+35}_{-26}$~\cite{Wang:2015vgv} &  Model-IV \\ \hline
			\multirow{3}{*}{$B \to \gamma$ form factor} & $365\pm 60$~\cite{Janowski:2021yvz} &  Model-I\\
			& $310\pm 60$~\cite{Janowski:2021yvz} &  Model-II\\
			& $415\pm 60$~\cite{Janowski:2021yvz} &  Model-III\\ \hline
			\multirow{2}{*}{$B \to \rho$ form factors} & $343^{+64}_{-79}$~\cite{Gao:2019lta} & Exponential Model  \\
			& $370^{+69}_{-86}$~\cite{Gao:2019lta} & Local Duality Model \\ \hline
			\hline \hline
		\end{tabular}
		\caption{\small A summary of $\lambda_B$ values from direct computation in QCD sum rule and indirect extractions using $B\to \pi$ and $B \to \gamma$ form factors (cf. the corresponding references for the details of models for the $B$-meson LCDA).}
		\label{tab:values}
	\end{table}

	In these circumstances, a complementary approach is vital and we make use of the recent developments in Lattice QCD form factor calculations which are generally available in the phase space region where the final state hadron has a low recoil momentum. In Ref. \cite{Parrott:2022rgu}, the form factors for the $B\to K \ell \ell$ decay are obtained across the full physical range of momentum transfer using the highly
	improved staggered quark formalism for all valence quarks on eight ensembles of gluon field configurations. The use of finer lattices makes it possible to generate data points in the kinematic region close to $q^2$ = 0, allowing larger momenta to be imparted to the daughter meson. 
    There are some previous analyses where predictions based on $z$-expansions are also given for the full range of $q^2$ by FNAL/MILC \cite{Bailey:2015dka} and HPQCD \cite{Bouchard:2013eph} collaborations for this channel. 
	
	In this analysis, we incorporate the values of the form factors obtained in the Ref. \cite{Parrott:2022rgu} at $q^2$ = 0 and express them in the framework of $B$-meson LCSRs, employing the two-particle LCDAs. We obtain the $1\,\sigma$ CI of the parameter $\lambda_B$. In section \ref{sec:theory}, we discuss the theoretical background, in section \ref{analysis}, we present the analysis and results and in section \ref{conc}, we conclude.
\section{Theoretical background} \label{sec:theory}

	In this section we review the theoretical framework for the LCSRs for $B$-meson to pseudoscalar meson transition form factors using $B$-meson LCDAs. This is a well-known approach~\cite{Faller:2008tr} in which the set of $B$-meson LCDA serves as a universal non-perturbative input and has been widely implemented for several transitions of $B$-meson. The starting point of this method is the $B$-to-vacuum correlator
	\begin{align}
	\label{eq:correlator}
	\!\!\!\mathcal{F}^{\mu\nu}(q, k)
	= i\! \!\int\! \text{d}^4 x\, e^{i k\cdot x}\,
	\bra{0} \mathcal{T}\lbrace J_{int}^{\nu}(x), J_{weak}^{\mu}(0)\rbrace \ket{\bar{B}_{q_2}(q + k)},
	\end{align}
	of two quark currents. $J_{int}^{\nu} \equiv \bar{q}_2(x) \gamma^\nu \gamma_5 q_1(x)$ interpolates the pseudoscalar meson $P$ with the momentum $k$ and $J_{weak}^{\mu}(0) \equiv \bar{q}_1(0) \Gamma_w^\mu b(0)$ is the weak current with the momentum $q$ related to $B\to P$ transition. The $B$-meson momentum being on-shell, $p^2\equiv (q+k)^2=m_B^2$. The correlator in Eq. \eqref{eq:correlator} is related to the form factor of our interest via the dispersion relation as
	\begin{equation}
	\label{eq:disp}
	\mathcal{F}^{\mu\nu}_{\rm had}(q,k)=\frac{\left\langle 0\left|\bar{q}_2 \gamma^\nu\gamma_5 q_1\right| P(k)\right\rangle\left\langle P(k)\left|\bar{q}_1 \Gamma^\mu_w b\right| \bar{B}(q+k)\right\rangle}{m_P^2-k^2}+\ldots\,,
	\end{equation}
	where the ellipses denote the higher excited and continuum states. The first term in Eq.~\eqref{eq:disp} is cast in terms of the pesudoscalar meson decay constant
	\begin{equation}
	\left\langle 0\left|\bar{q}_2 \gamma^\nu \gamma_5 q_1\right| P(k)\right\rangle=i k^\nu f_P\,.
	\end{equation}
	The second term denotes $B\to P$ hadronic matrix element for the weak current. In the case of vector and tensor weak current i.e., $\Gamma^\mu_w=\gamma^\mu$ and $\sigma^{\mu \nu}q_\nu$, respectively, the hadronic matrix element can be parameterized using the three independent non-vanishing form factors $f_+$, $f_0$ and $f_T$,
	\begin{align}
		\label{eq:BtoP:vector}
		\bra{P(k)} \bar{q}_1 \gamma^\mu b \ket{B(p)}
		& = \left[(p + k)^\mu - \frac{m_B^2 - m_P^2}{q^2}q^\mu\right]\, f_+
		+ \frac{m_B^2 - m_P^2}{q^2} q^\mu\, f_0,\\
		\label{eq:BtoP:tensor}
		\bra{P(k)} \bar{q}_1 \sigma^{\mu\nu}\, q_\nu b \ket{B(p)}
		& = \frac{i f_T}{m_B + m_P}\, \left[q^2\, (p + k)^\mu - (m_B^2 - m_P^2)\, q^\mu\right]\,.
	\end{align}
	The form factors are functions of the momentum transfer $q^2 \equiv (p - k)^2$, where
	$p$ and $k$ denote the $B$-meson and the final-state meson momenta, respectively, and we drop the explicit dependence for notational simplicity.

	The LCSRs for the form factors can then be obtained by matching the hadronic representation in Eq. \eqref{eq:disp} to the OPE calculation using semi-global quark-hadron duality. For the computation of the OPE part, the correlator Eq.~\eqref{eq:correlator} is expanded at near light-cone separations $x^2 \simeq 0 $, implying $k^2 \ll m_{q_1}^2$ and $q^2 \ll (m_b+m_{q_1})^2$ in the momentum space. 
	Neglecting the gluon emission effects from the intermediate $q_1$-quark propagator, the correlators can be written as
	\begin{align}
		\label{eq:corrOPE}
		{\mathcal F}_{\rm OPE}^{\mu\nu}( q,k) = \!\int \!d^4 x\, e^{ik\cdot x}\!\!\int \frac{d^4l}{(2\pi)^4}e^{-il\cdot x}
		\big[\gamma^\nu \gamma_5
		\frac{\slashed{l}+m_{q_1}}{m_{q_1}^2-l^2}
		\Gamma^\mu_w \big]_{\alpha\beta}
		\langle 0|\bar{q}_2^\alpha(x)b^\beta(0)  | \bar{B}(q+k)\rangle
		\,,  
	\end{align}
	where $\alpha,\beta$ are spinor indices.
	The non-local $B$-to-vacuum matrix element $\left\langle 0\left|\bar{q}_2^\alpha(x) b^\beta(0)\right| \bar{B}(q+k)\right\rangle$ then can be approximated, in the heavy quark effective theory limit, in terms of $B$-meson LCDAs of increasing twist (see \ref{app:BDA} for the expression). The result for the correlation function is equated to the hadronic representation
	(in Eq.~\eqref{eq:disp}) with semi-global quark-hadron duality assumption for which the Borel transformation from $k^2 \to M^2$ is used which exponentially suppresses the contributions from excited resonances and continuum states. Finally, each independent Lorentz-structure in this equation provides a sum rule relation for a certain form factor or a combination of form factors which then can be written in the following compact form for all the form factors $F$ \cite{Gubernari:2018wyi},
	\begin{align} \label{eq:FF}
		F= &\frac{f_B M_B}{K^{(F)}} \sum_{n=1}^{\infty}  
		\Big{\{}(-1)^{n}\int_{0}^{\sigma_0} d \sigma \;e^{(-s(\sigma,q^2)+m_{P}^2)/M^2} \frac{1}{(n-1)!(M^2)^{n-1}} I_n^{(F)}  \nonumber \\
		& - \Big{[}\frac{(-1)^{n-1}}{(n-1)!} e^{(-s(\sigma,q^2)+m_{P}^2)/M^2} \sum_{j=1}^{n-1} \frac{1}{(M^2)^{n-j-1}}\frac{1}{s'} (\frac{d}{d \sigma} \frac{1}{s'})^{j-1} I_n^{(F)} \Big{]}_{\sigma=\sigma_0} \Big{\}}\,,
	\end{align}
	
	where 
		\begin{equation}
		\sigma = \frac{\omega}{m_B},~~~~ s(\sigma,q^2) = \sigma m_B^2 + \frac{m_1^2 -\sigma q^2}{\bar{\sigma}},~~~~ s'(\sigma,q^2) = \frac{d s(\sigma,q^2)}{d \sigma}, ~~~~ \bar{\sigma} = 1-\sigma\,.
	\end{equation}
 
	Here, $\sigma_0$ = $\sigma(s_0,q^2)$ where $s_0$ is the effective threshold which is expected to be close to the mass square of the first excited state.
	The effective threshold can either be calculated using the two-point sum rule for the decay constant of the pseudoscalar meson and then be used in the LCSRs (see Ref.~\cite{Gubernari:2022hrq} as an example of such analysis for $B \to D_1$ form factor) or can also be obtained from the LCSR~\cite{Gubernari:2018wyi}. The parameter $M^2$ is the Borel parameter. The choice of Borel parameter interval is made such that the impact of higher-twist contributions, which varies as powers of 1/$M^2$, are suppressed. Moreover, the validity of the quark-hadron duality approximation has to be ensured with the exponential suppression of the higher excited states above $s_0$ by the choice of $M^2$.

	As mentioned earlier, in this analysis, we have considered the two-particle LCDAs defined through the functions $I_n^{(F)}$:
	
	\begin{equation}
	I_n^{(F,2p)}(\sigma,q^2) = \frac{1}{\bar{\sigma}^n} \sum_{\psi_{2 p}} C_n^{(F,\psi_{2p})}(\sigma,q^2) \psi_{2 p} (\sigma m_B),
	\end{equation}
	where $\psi_{2 p} = \phi_+, \bar{\phi}, g_+, \bar{g}$ with
	\begin{subequations}
		\begin{align}
		&\bar{\phi}(w) = \int_0^w d \eta (\phi_+ (\eta) - \phi_-(\eta)) \,, \\
		&\bar{g}(w) = \int_0^w d \eta (g_+(\eta) - g_-(\eta))\,.
		\end{align}
	\end{subequations}
	
	The LCDAs $\phi_+, \phi_-, g_+$ and $g_-$ are of twist two, three, four, and five, respectively. In this work, we have considered the parametrization of the $B$-meson LCDAs in the Exponential and Local Duality models \cite{Braun:2017liq} in order to check the model dependence of the parameter $\lambda_B$. The Exponential Model is based on combining the regime of low momentum of quarks and gluons with an exponential suppression at large momentum, whereas the Local Duality Model is based on the duality assumption to match the $B$-meson state with the perturbative spectral density integrated over the duality region. For the latter, all two-particle $B$-meson LCDAs vary as $\sim$ $(2w_0-w)^p$. In this work, we have considered the simplest case of $p = 1$. 
	The expressions for these models are compiled in \ref{app:BDA}. The main input involved in parameterizing the LCDAs is the first inverse moment of twist two LCDA defined as
	\begin{equation}
	\lambda_B = \Big{(}\int_0^{\infty} dw \frac{\phi_+(w)}{w} \Big{)}^{-1}\,.
	\end{equation}
	Apart from the parameter $\lambda_B$, these LCDAs also depend on the parameters $\lambda_E$ and $\lambda_H$ which correspond to matrix elements representing quark-gluon three-body components in the $B$-meson wavefunction.
	
	In this analysis, the input values for $\lambda_E^2$ and $\lambda_H^2$ are taken from Ref. \cite{Nishikawa:2011qk} and the $B$-meson decay constant $f_B$ from Ref.~\cite{FlavourLatticeAveragingGroupFLAG:2021npn} which corresponds to an average of the inputs from \cite{Bazavov:2017lyh,ETM:2016nbo,Dowdall:2013tga,Hughes:2017spc} as listed below.
	\begin{align} 
	\label{eq:fBval}
	f_B = & 190.0 \pm 1.3\, \text{MeV}\,,\qquad m_B = 5279.66 \pm 0.12\, \text{MeV}\,, \\ \label{eq:lambdaval}
	\lambda_E^2 = & 0.03 \pm 0.02 \,\text{GeV}^2\,, \qquad
	\lambda_H^2 =  0.06 \pm 0.03\, \text{GeV}^2 \,.
	\end{align}
	
	The coefficients $C^{(F,\psi)}$ and the normalization factors $K^{(F)}$ for the various form factors are obtained from Ref~\cite{Gubernari:2018wyi} and also been collected in \ref{app:BDB} for the reader’s convenience.
	
	\section{Analysis and results} \label{analysis}

	Using the theoretical expressions discussed in the previous section, we are now at a stage to perform the numerical analysis to extract the value for $\lambda_B$ using the form factor results from Lattice QCD approach. We notice that apart from the mass and decay constants of the mesons, there are several other input parameters which enter into the form factor expressions (in Eq.~\eqref{eq:FF}).  In this regard, we first define an
	optimized $\chi^2$- statistic as
	\begin{equation}
	\label{eq:chisq}
	\chi^2 = \sum_{i,j} (O_i^{\rm Lattice}- O_i^{\rm theo}).\, Cov^{-1}_{ij}. \,(O_j^{\rm Lattice}- O_j^{\rm theo}) + \chi^2_{\rm nuis}\,.
	\end{equation} 
	Here, $O_i^{\rm Lattice}$ corresponds to the values of the form factors $f_+$ and $f_T$ for the $B \to K$ transition at $q^2$ = 0\footnote{Note that at $q^2$ = 0, as $f_+(0)$ = $f_0(0)$, we do not include $f_0(0)$ in the $\chi^2$ statistic.} provided by the HPQCD collaboration \cite{Parrott:2022rgu} and $Cov$ is the corresponding covariance matrix which includes the correlation between these form factors. We have assigned $\chi^2$ corresponding to the nuisance parameters $\lambda_E^2$, $\lambda_H^2$, $s_0$ and $M^2$
	in $\chi^2_{\rm nuis}$ where the parameters $\lambda_E^2$, $\lambda_H^2$ and  $s_0$ follow Gaussian distributions, whereas $M^2$ follows a uniform distribution, with the corresponding 1$\sigma$ CIs. quoted in Table \ref{tab:Inputs} and Eq. \eqref{eq:lambdaval}.
	
	We obtain $\lambda_B$ = 0.34 $\text{GeV}$ (0.47 $\text{GeV}$) as the best fit estimate while using the Exponential Model (Local Duality Model) for the $B$-meson LCDAs entering in the $B \to K$ form factors. Here we find $\chi^2_{min}/$dof = $1.04/1$ in the Exponential Model and $\chi^2_{min}/$dof = $0.35/1$ in the Local Duality Model. In order to obtain the 1$\sigma$ CI of the fit parameter $\lambda_B$, we vary $\lambda_B$ within the range\footnote{This range has been chosen so that it contains the best-fit estimate.} (0.1,\,0.9) GeV where for each value of $\lambda_B$, we minimize the $\chi^2$ defined in Eq. \eqref{eq:chisq} over the range of nuisance parameters. This variation results into different larger values of $\chi^2$ around $\chi^2_{min}$. At the best fit point, $\Delta \chi^2$ = 0 and the 1\,$\sigma$ or 68 $\%$ confidence interval corresponds to the region bounded by the $\Delta \chi^2$ = 1 curve. This gives us the uncertainty associated with $\lambda_B$. Our results are 
	\begin{align}
	\lambda_B (1\,\text{GeV}) &= 338^{+ 68}_{-9}\, \text{MeV} ~~~~ (\text{Exponential Model})\,,\label{eq:lambdaBValue}\\
     \lambda_B (1\,\text{GeV})&= 472^{+110}_{-41} \text{MeV} ~~~(\text{Local Duality Model})\,.
	\label{eq:lambdaBValue1}    
	\end{align}

\begin{table*}[t]
	\centering
	\renewcommand*{\arraystretch}{1.3}
	\begin{tabular}{|c|c|c|c|}
		\hline
		Meson & Decay constant $(f_P)$ [MeV] & $s_0$ $[\text{GeV}^2]$ & $M^2$ $[\text{GeV}^2]$   \\
		\hline\hline
		
		$\pi$ & 130.2 $\pm$ 0.8 \cite{FlavourLatticeAveragingGroupFLAG:2021npn,RBC:2014ntl,Follana:2007uv,MILC:2010hzw} & 0.7 $\pm$ 0.014 \cite{Khodjamirian:2003xk,Colangelo:2000dp} & 1.0 $\pm$ 0.5 \cite{Khodjamirian:2006st,Faller:2008tr} \\
		$K$ & 155.7 $\pm$ 0.3 \cite{FlavourLatticeAveragingGroupFLAG:2021npn,FermilabLattice:2014tsy,Dowdall:2013rya,Carrasco:2014poa} & 1.05 $\pm$ 0.021 \cite{Khodjamirian:2003xk,Colangelo:2000dp} & 1.0 $\pm$ 0.5 \cite{Khodjamirian:2006st,Faller:2008tr} \\
		$D$ & 212.0 $\pm$ 0.7 \cite{FlavourLatticeAveragingGroupFLAG:2021npn,Carrasco:2014poa,Bazavov:2017lyh} & [5.8,7.8] \cite{Gubernari:2018wyi} & 4.5 $\pm$ 1.5 \cite{Khodjamirian:2006st,Faller:2008tr} \\
		
		\hline\hline
	\end{tabular}
	\caption{\small Inputs used in this analysis.}
	\label{tab:Inputs}
\end{table*}

	We notice that the value of $\lambda_B$ is influenced by the choice of $B$-meson LCDA model, yet both the values are in agreement with the previous QCD sum rule estimates within the $\pm1\sigma$ uncertainty quoted in Table.~\ref{tab:values}. However note that the uncertainty in our result is reduced almost by a factor of two especially for the case with Exponential Model. The difference between the results obtained in the two LCDA models can probably be thought of as systematic uncertainty associated with the lack of complete knowledge or unique choice of LCDA parametrization.
	The contribution to the $B\to K$ form factors arising from the gluon-emission from $s$-quark is found to be less than 1\%~\cite{Gubernari:2018wyi}. This justifies the fact that such effects can be safely neglected. In the above indirect extractions, a `reasonable' choice of the factorization scale $\mu =1\,\text{GeV}$ is assumed at which the values of other input parameters entering in the sum rule are considered\footnote{The renormalisation scale dependence of $B$-meson LCDAs can be found in~\cite{Lange:2003ff}.}. Observing the asymmetric errors associated with the estimation of $\lambda_B$, we also quote the 2 $\sigma$ CI as $\lambda_B = 338^{+ 89}_{-21}\, \text{MeV} (\text{Exponential Model})$ and $\lambda_B = 472^{+139}_{-69} \text{MeV} (\text{Local Duality Model})$. However, from a conservative viewpoint, in the subsequent analyses, we have symmetrised the upper limit of the uncertainty.   
	
	\begin{figure*}
		\centering
		\includegraphics[width=0.32\linewidth]{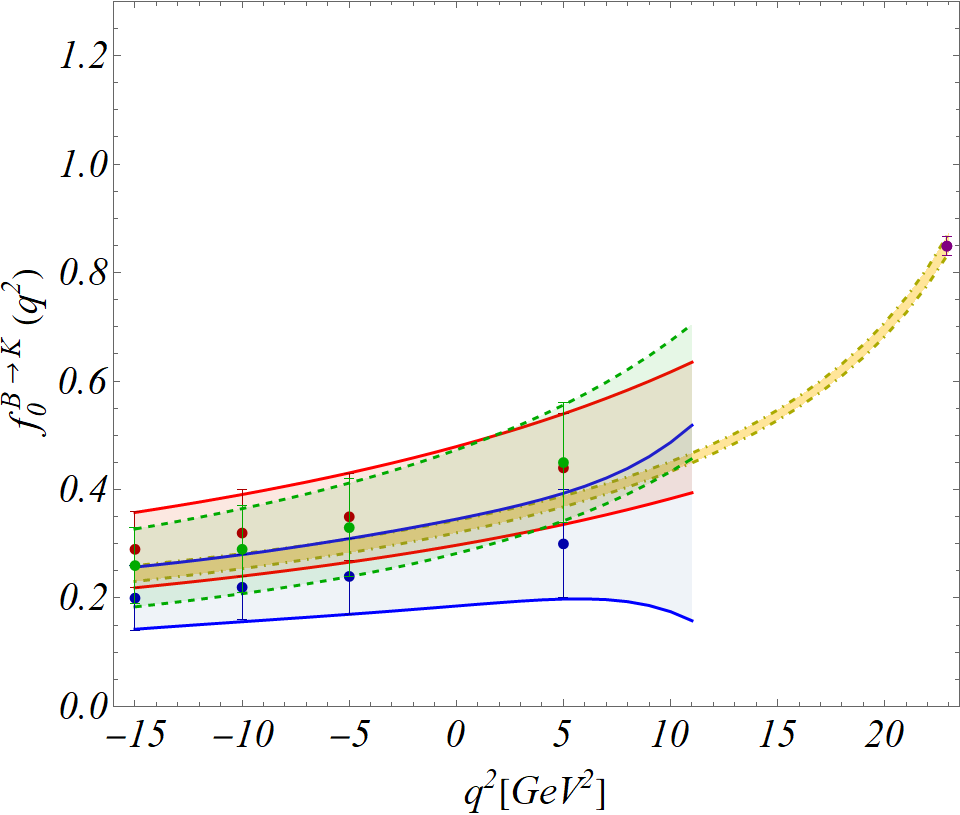}~~~
		\includegraphics[width=0.32\linewidth]{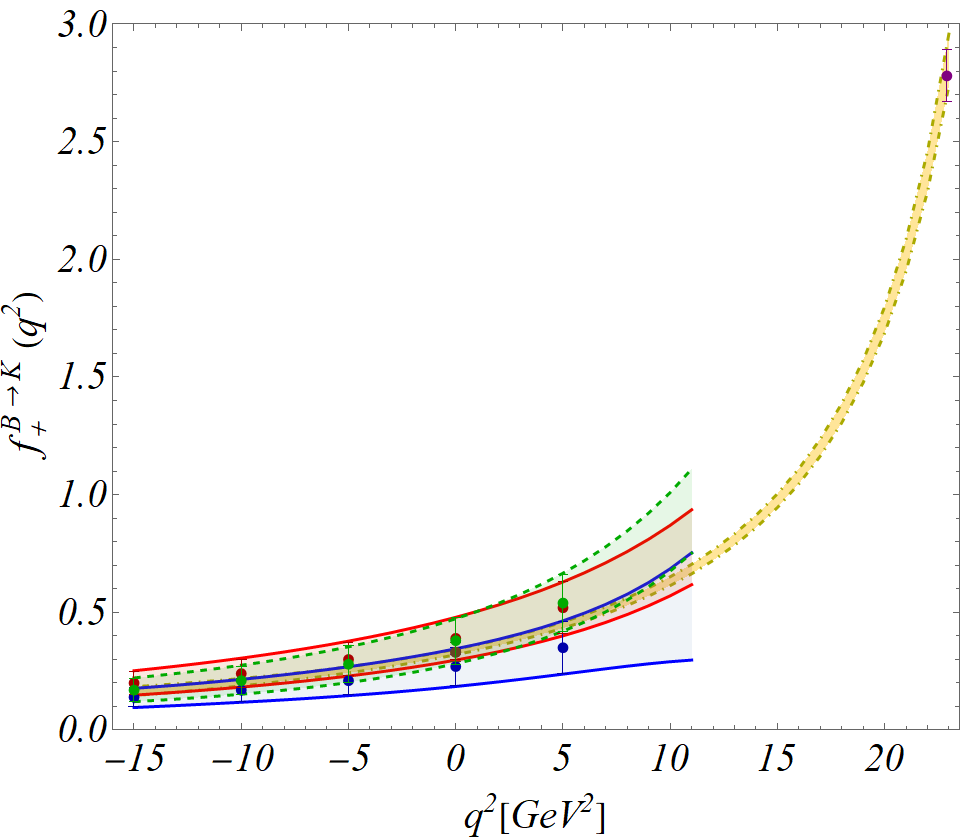}~~~
		\includegraphics[width=0.32\linewidth]{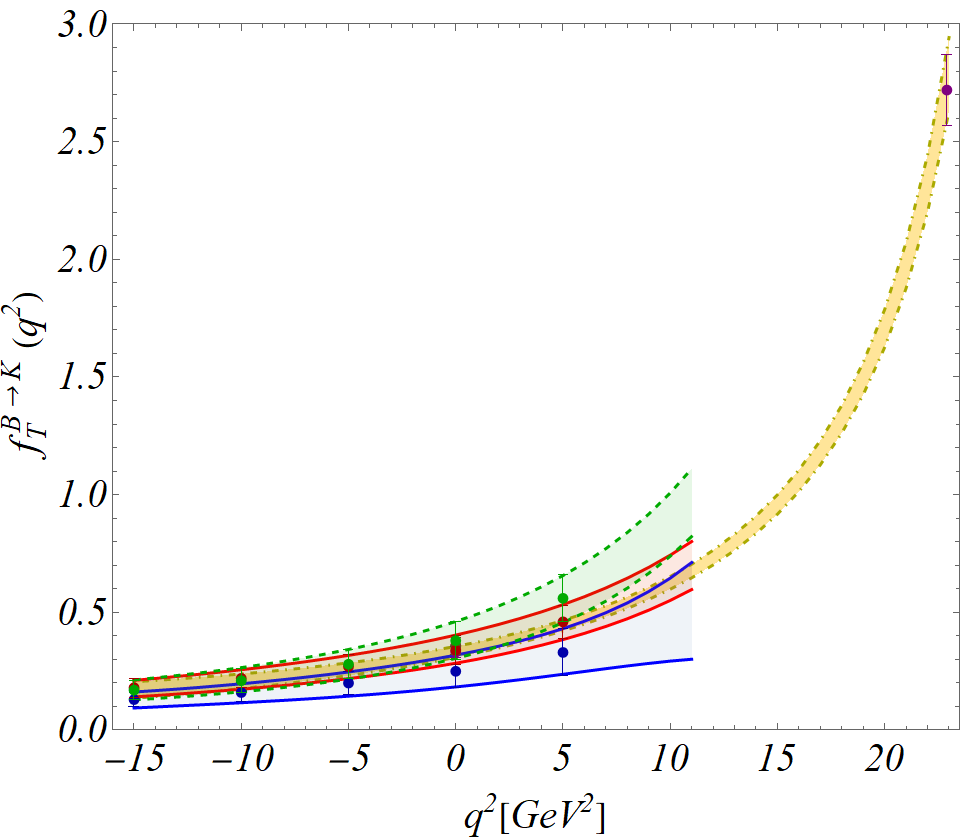}
		\includegraphics[width=0.47\linewidth]{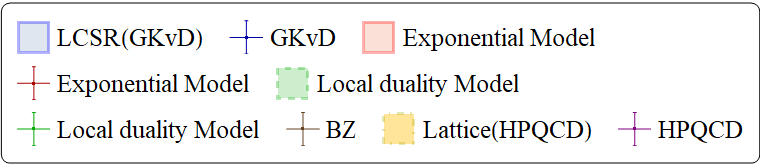}
		\caption{\small Central values and 68\% confidence interval for the variation of the three form factors, $f_0,~f_+,~f_T$, for $B \to K$ channel, with the momentum transfer $q^2$. The red (green) band and error bars are the LCSR prediction based on our result of $\lambda_B$ (Eqs.~\eqref{eq:lambdaBValue} and \eqref{eq:lambdaBValue1}) obtained using the Exponential (Local Duality) Model. The blue and yellow bands correspond to LCSR estimates from GKvD\cite{Gubernari:2018wyi} and Lattice results from HPQCD~\cite{Parrott:2022rgu} collaboration, respectively. The brown error bar represents the earlier LCSR calculation from BZ \cite{Ball:2004ye}, utilizing $K$-meson LCDAs and the purple error bar denotes the Lattice estimates from HPQCD \cite{Parrott:2022rgu}. }
		\label{fig:B2Kcomp}
	\end{figure*}
	
\begin{figure*}
		\small
		\centering
		\includegraphics[width=0.37\textwidth]{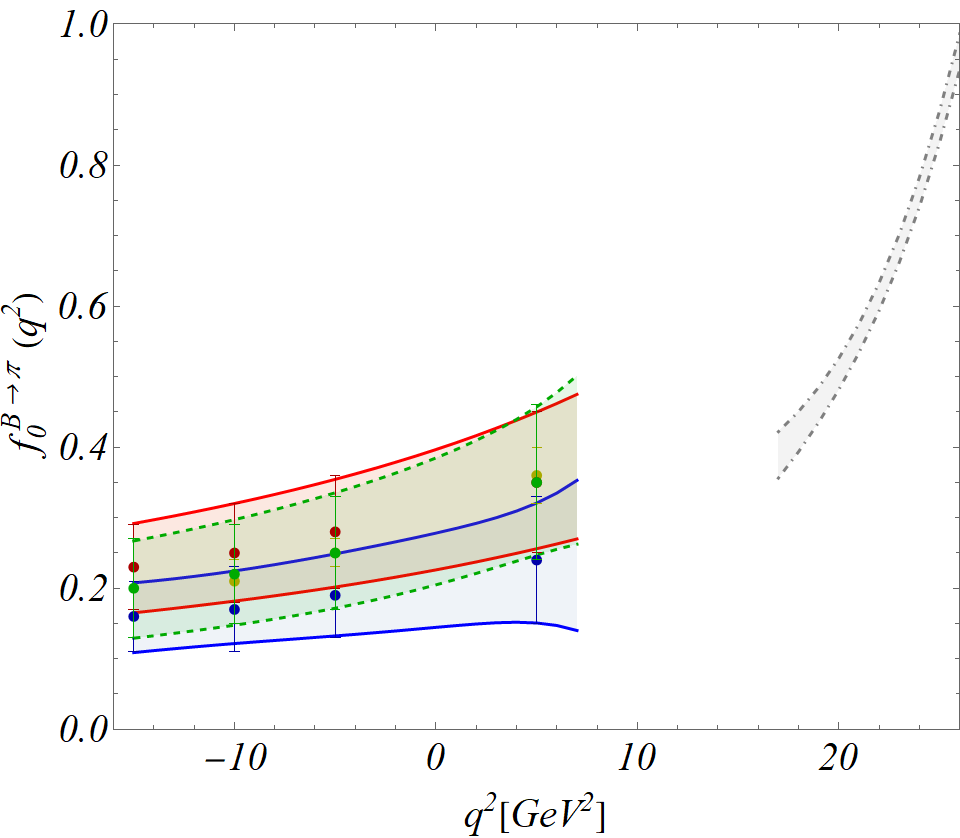}%\label{fig:f0B2Pi}
		~~~~~~~
\includegraphics[width=0.37\textwidth]{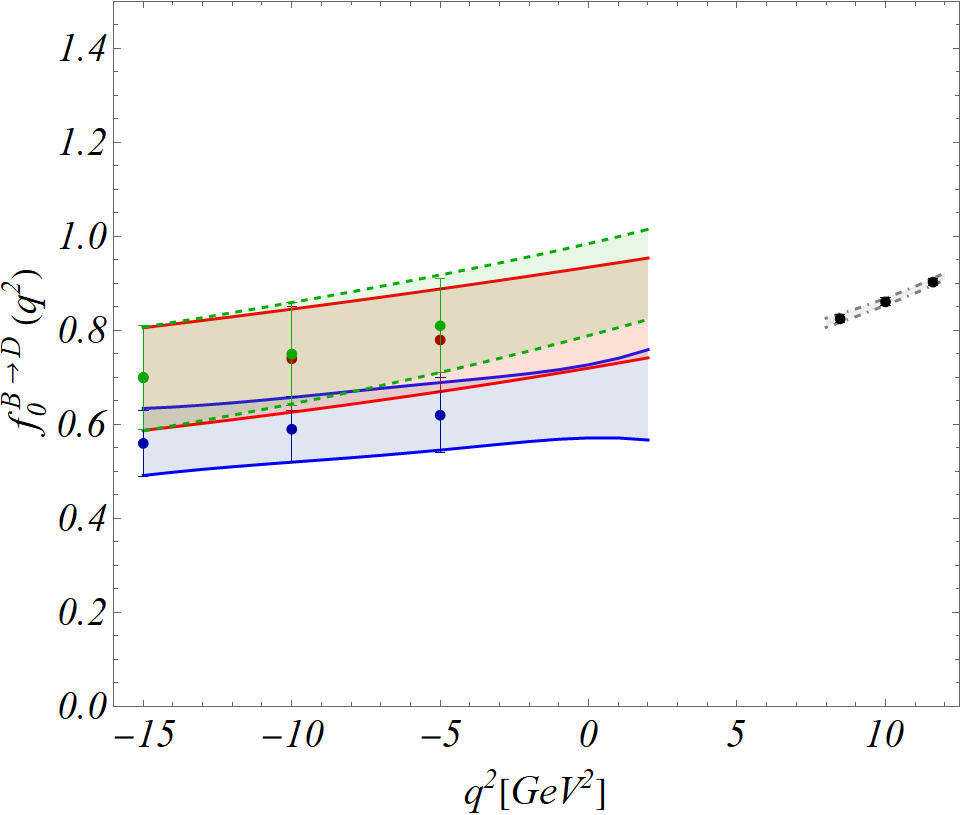}%\label{fig:f0B2D}
		\\[2ex]
  \includegraphics[width=0.37\textwidth]{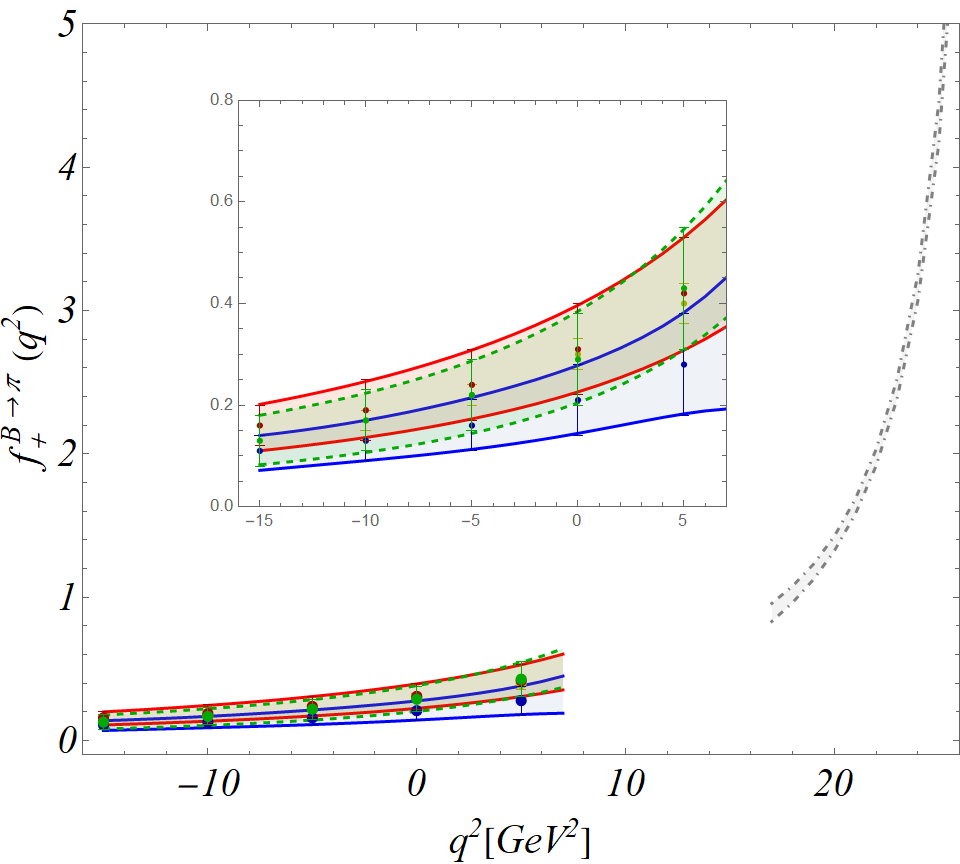}%\label{fig:fplusB2Pi}
~~~~~~~	\includegraphics[width=0.37\textwidth]{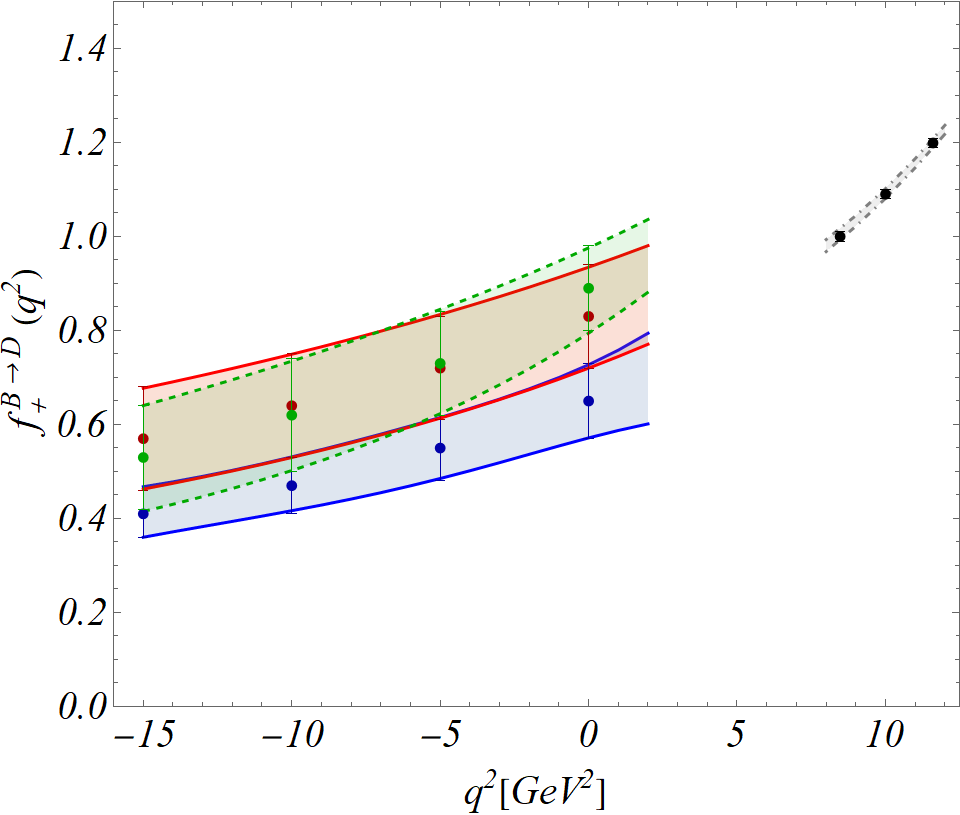}%\label{fig:fplusB2D}
  \\[2ex]
  \includegraphics[width=0.37\textwidth]{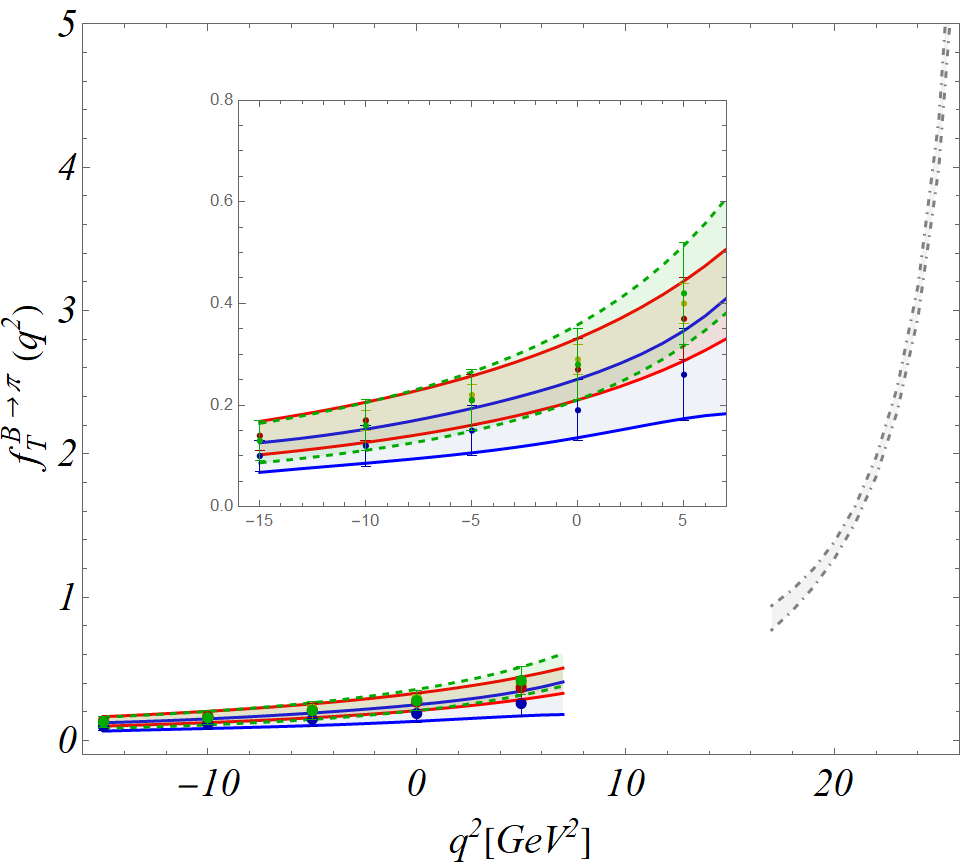}%\label{fig:fTB2Pi}
		~~~~~~~
		\includegraphics[width=0.37\textwidth]{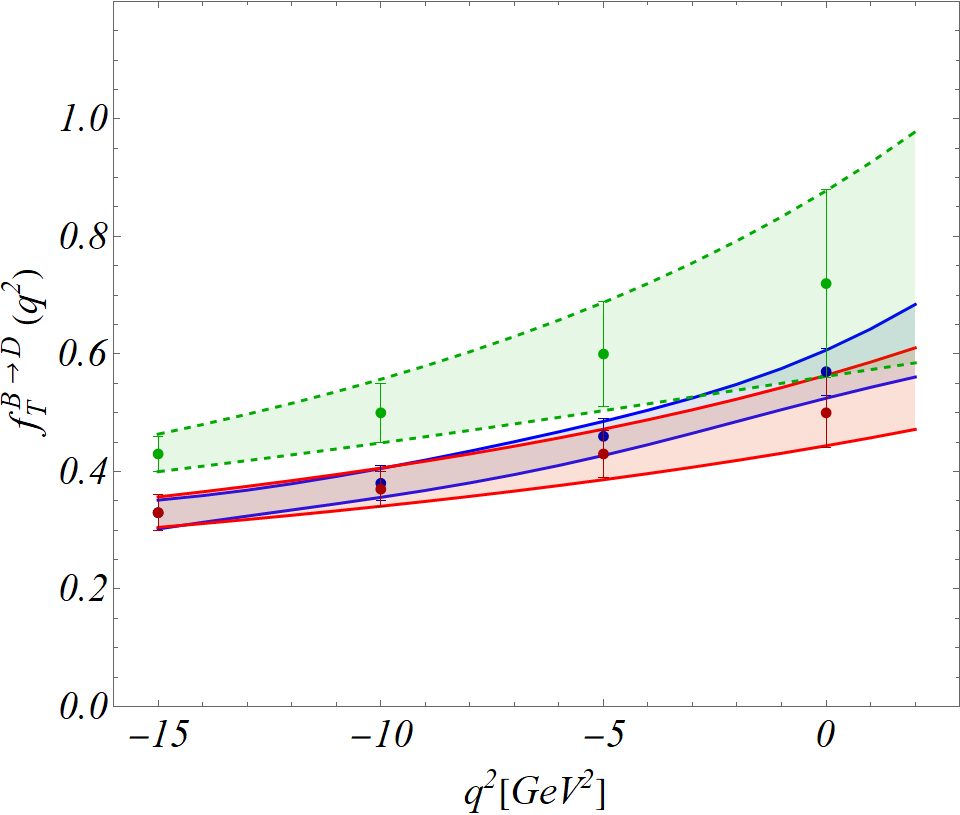}%\label{fig:fTB2D}
		\\
		\includegraphics[width=0.35\textwidth]{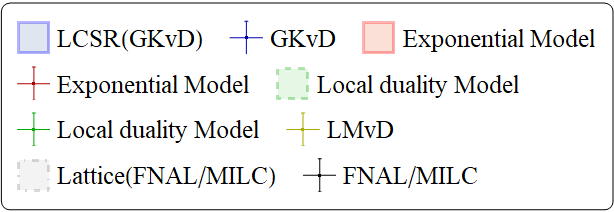}%\label{fig:legend}
		\caption{\small Variation of the form factors with momentum transfer $q^2$ for $B \to \pi,\,D$ modes. The left(right) column shows the plots for $B \to \pi(D)$ channels. The red (green) band and error bars represents LCSR predictions using our extracted values of $\lambda_B$ (Eqs.~\eqref{eq:lambdaBValue} and \eqref{eq:lambdaBValue1}). The blue bands and error bars correspond to LCSR estimates from GKvD\cite{Gubernari:2018wyi}. The grey bands and black error bars correspond to Lattice estimates from FNAL/MILC collaboration (see Ref. \cite{MILC:2015uhg} for $B \to D$ and Refs. \cite{FermilabLattice:2015mwy,FermilabLattice:2015cdh} for $B \to \pi$ decays). For the $B \to \pi$ case, the yellow error bars denote LCSR predictions using $\pi$-meson LCDAs from LMvD~\cite{Leljak:2021vte}, and the insets for form factors $f_+$ and $f_T$ show the magnified part of low $q^2$ region for better visualization.  }
		\label{fig:plots}
	\end{figure*}

	It would now be interesting to compare the predictions for the form factors obtained in this work for $B \to P~(P = K,\pi,D)$ channels with the already existing analyses using our estimates of $\lambda_B$ (from Eqs.~\eqref{eq:lambdaBValue} and \eqref{eq:lambdaBValue1}) and with the corresponding $B$-meson LCDA parametrization. In this regard, using Eq.~\eqref{eq:FF} we have created LCSR data points for the form factors $f_+$ and $f_T$ at $q^2 =\{ -15,\,-10,\,-5,\,0,\,5\}\,\text{GeV}^2$ and for $f_0$ at $q^2 = \{-15,\,-10,\,-5,\,5\}\,\text{GeV}^2$ for the modes $B \to K$ and $B \to \pi$. However, for the $B \to D$ mode, we do not go beyond $q^2$ = 0 $\text{GeV}^2$ because for positive values of $q^2$, the relative contribution from the higher-twist two-particle terms increases, which in turn makes the calculation of $B \to D$ form factors unstable (in this region), as mentioned in Ref. \cite{Gubernari:2018wyi}. The ranges for the parameters $\lambda_E^2$, $\lambda_H^2$, $s_0$ and $M^2$ as provided in Table \ref{tab:Inputs} and Eq.~\eqref{eq:lambdaval}, and the extracted values of $\lambda_B$ from Eqs. \eqref{eq:lambdaBValue} and \eqref{eq:lambdaBValue1} have been used here.
	In Figs. \ref{fig:B2Kcomp} and \ref{fig:plots}, we have shown the variation of the form factors $f_+$, $f_0$ and $f_T$ with $q^2$ for $B \to K$, $B \to \pi$ and $B\to D$ channels. We use $z$-parametrization for the extrapolation of form factors from negative $q^2$ to the physical range where, in this analysis, we include the correlation between different form factors for the same mode. The relevant expressions and the expansion coefficients obtained are given in \ref{app:zfit}. 

	For the $B \to K$ mode, we compare our results obtained within the Exponential Model (red bands and error bars) and the Local Duality Model (green bands and error bars) with those from Ref. \cite{Gubernari:2018wyi}[GKvD] (blue bands and error bars). In the latter reference, form factors are computed within the LCSR approach using the $B$-meson LCDA (Exponential Model), while employing the QCD sum rule estimate of $\lambda_B$~\cite{Braun:2003wx} as input. Additionally, we compare our findings to the most recent Lattice estimates from the HPQCD collaboration~\cite{Parrott:2022rgu} (yellow bands and purple error bars) and the earlier LCSR calculation presented in Ref. \cite{Ball:2004ye}[BZ] (brown error bars), where $K$-meson LCDAs were utilized.
	As anticipated, our results align well with the Lattice estimates at $q^2$ = 0 $\text{GeV}^2$ (this point is difficult to distinguish from other colours in the plot). Notably, we observe that the central values of the form factors at various $q^2$ points, as obtained in our case for both models, tend to be relatively higher than those reported in Ref. \cite{Gubernari:2018wyi}. However, these values remain within the $\pm1\sigma$ uncertainty range, except for the form factor $f_T$ in the Local Duality Model for $q^2 \gtrsim 5 \,\text{GeV}^2$. Remarkably, our results for all the form factors, derived in both models, exhibit a high degree of consistency with each other, as well as with the findings of Ref. \cite{Ball:2004ye}, which are specifically available at $q^2 = 0 \,\text{GeV}^2$. 
	
 	For the $B \to \pi$ case, we compare our results with those from Ref. \cite{Leljak:2021vte}[LMvD] (represented by yellow error bars). In this reference, form factors are computed using LCSR methods with $\pi$-meson LCDAs. We also compare our findings with those from Ref. \cite{Gubernari:2018wyi}[GKvD] (depicted by blue bands and error bars), which employs the $B$-meson LCDAs as previously mentioned. We have also shown Lattice estimates from the FNAL/MILC collaboration \cite{FermilabLattice:2015mwy,FermilabLattice:2015cdh} in grey bands for $q^2$ $\gtrsim$ 17 $\text{GeV}^2$ since the lattice simulations are most precise in this region. Note that the analyses provide only the fit results for the coefficients of the $z$-expansions of the respective form factors. It is noteworthy that our results exhibit excellent agreement with those of [LMvD] in the Exponential Model (depicted by red bands and error bars) for $f_T$ at $q^2 = -10 \,\text{GeV}^2$, as well as in the Local Duality Model (represented by green bands and error bars) for $f_0$ at $q^2 = -5 \,\text{GeV}^2$ and for $f_+$ at $q^2 = -10,\, -5 \,\text{GeV}^2$.
	Moreover, our results for all form factors, obtained in both models, remain consistent with those of [GKvD] and [LMvD]. However, it is worth noting that there is a disagreement between their results for $f_T$ at $q^2= 0,\, 5 \,\text{GeV}^2$ by 1.5 $\sigma$ and 1.4 $\sigma$, respectively.
	
	For the $B\to D$ channel, similar comparison is made with Ref. \cite{Gubernari:2018wyi}[GKvD], and we infer from Fig. \ref{fig:plots} (lower panel), that our estimates of the form factors $f_+$ and $f_0$ obtained in both the models are consistent with those of [GKvD]. However, it is worth noting that the results based on the  Local Duality Model for $f_T$ are not consistent (within $\pm 1\sigma$ range) with  those of [GKvD] and also with our prediction in Exponential Model (Eq.~\eqref{eq:lambdaBValue}) for $q^2 \lsim -5 \,\text{GeV}^2$. This mismatch might be attributed to the unknown systematic uncertainties associated with the model dependence of $\lambda_B$, as also discussed before. For completeness, we have also shown the lattice estimates from the FNAL/MILC collaboration \cite{MILC:2015uhg}in grey bands and black error bars for $f_+$ and $f_0$ form factors for $q^2$ $\gtrsim$ 8 $\text{GeV}^2$.
	
	With our extracted values of $\lambda_B$ (Eqs.~\eqref{eq:lambdaBValue} and \eqref{eq:lambdaBValue1}), we proceed to calculate the branching fraction of the radiative mode $B \to \gamma \ell \nu$. This branching fraction, in addition to $\lambda_B$, is influenced by the logarithmic moments of the leading-twist $B$-meson LCDA $\phi_+$ \cite{Janowski:2021yvz, Beneke:2018wjp}. The obtained branching fraction is parameterized as a function of $\lambda_B$ for various minimum photon energy thresholds: $E_{\gamma} > 1,\,1.5,\,2\, \text{GeV}$. This behavior is illustrated in Fig.~9 of Ref. \cite{Beneke:2018wjp}, where the uncertainty band is determined by varying the logarithmic moments within the specified ranges.
	
	By overlaying the $1\,\sigma$ CI of $\lambda_B$ obtained in this work for both models onto Fig.~9 of Ref. \cite{Beneke:2018wjp}, we predict the corresponding $1\,\sigma$ CI for the branching fraction in this mode. 
	Specifically, for the Exponential Model, we get
	\begin{equation} \label{eq:radexp}
	{\mathcal B}(B \to \gamma \ell \nu ) =\! 
	\left\{\hspace{-1mm}
	\begin{array}{lcl}
	\displaystyle (0.89 \pm 0.10) \times 10^{-6} &\hspace{-5mm} & (E_{\gamma}\! > \!1.5\,\text{GeV}),
	\\[2ex]
	(0.34 \pm 0.02) \times 10^{-6}  &\hspace{-5mm} & (E_{\gamma} \!>\! 2.0\, \text{GeV})\, ,
	\end{array}
	\right. 
	\end{equation}
	and for the Local Duality Model,
	\begin{equation} \label{eq:radloc}
	{\mathcal B}(B \to \gamma \ell \nu ) = \!
	\left\{\hspace{-1mm}
	\begin{array}{lcl}
	\displaystyle (0.36 \pm 0.03) \times 10^{-6} &\hspace{-5mm} & (E_{\gamma}\! >\! 1.5 \,\text{GeV}),
	\\[2ex]
	(0.14 \pm 0.01) \times 10^{-6}  &\hspace{-5mm} & (E_{\gamma} \!>\! 2.0 \, \text{GeV})\, .
	\end{array}
	\right. 
	\end{equation}
	We notice that both the predicted branching fraction and its associated uncertainty decrease as the minimum photon energy cut is raised.
	These predictions satisfy the experimental upper limit of $3 \times 10^{-6}$ obtained at a 90$\%$ CI \cite{Belle:2018jqd}. As the branching fraction prediction for this channel depends strongly on the parameter $\lambda_B$, the values in Eqs. \eqref{eq:radexp} and \eqref{eq:radloc} deviate from each other by several sigma due to the different extracted values of $\lambda_B$ in the two LCDA models considered in this work.

	\section{Conclusion} \label{conc}
	
	In this analysis, we have constrained the first inverse moment of the $B$-meson light-cone distribution amplitude, $\lambda_B$ using the recent Lattice QCD results for the $B \to K$ form factors at zero momentum
	transfer ($q^2=0$) from the HPQCD collaboration. Our estimated values for $\lambda_B$ are found to be $\lambda_B$ = $338^{+ 68}_{-9}$ $\text{MeV}$ ($472^{+ 110}_{-41}$ $\text{MeV}$) using the Exponential Model (Local Duality Model) for the $B$-meson LCDA, exhibiting agreement with earlier QCD sum rule estimates within a $\pm 1 \sigma$ uncertainty. We observe a model dependence in $\lambda_B$. Furthermore, the uncertainty improvement achieved in this estimation with the Exponential Model is nearly a factor of two compared to previous results, when excluding systematic uncertainties due
to the stated model dependence.
	
	With the $\lambda_B$ values obtained in the two models, we have compared the form factor results for $B \to P~(P = K,\pi,D)$ channels with earlier analyses that utilized either light meson LCDAs or $B$-meson LCDAs with QCD sum rule estimate of $\lambda_B$. Our findings show a high level of consistency with those previous analyses, except for the form factor $f_T$ in the Local Duality Model for the $B \to K$ and $B \to D$ channels in specific $q^2$ regions. The constrained value of $\lambda_B$ obtained in this work using the well-accepted Exponential Model holds the potential to facilitate more precise predictions of $B$-meson transition form factors within the LCSR approach, employing $B$-meson LCDAs. Using these estimates, we have also determined the branching fraction for the $B \to \gamma \ell \nu$ mode, and we observe that they satisfy the experimental upper limit provided by the Belle collaboration.
	
	\subsection*{Acknowledgment}
	R.M. thanks Alexander Khodjamirian for the valuable discussion. The work of I.R. is supported by IIT Gandhinagar Research and Development Grant.
	
	\appendix
	\section{$B$-meson light-cone distribution amplitudes}
	\label{app:BDA}
	
	Following the definitions of the $B$-meson LCDAs provided in Ref.~\cite{Braun:2017liq}, we present the expression for the $B$-to-vacuum matrix element in the heavy quark effective theory limit. In this limit, the heavy $b$-quark field is substituted with the heavy quark effective theory field $h_v$, where $v^\mu=\left( q+k\right)^\mu/m_B$ represents the $B$-meson's four-velocity in its rest frame.
	The expansion in terms of two-particle $B$-meson LCDAs reads
		\begin{multline}
		\bra{0} \bar{q}^{\alpha}(x) h_{v}^{\beta}(0) \ket{\bar{B}(v)} =
		-\frac{i f_B m_B}{4} \int^\infty_0 d\omega  \bigg\{
		(1 + \slashed{v}) \bigg[
		\phi_+(\omega) -g_+(\omega) \partial_\lambda \partial^\lambda
		\\*
		+\frac12 \left(\overline{\Phi}_{\pm}(\omega)
		-\overline{G}_\pm(\omega) \partial_\lambda \partial^\lambda\right) 
		\gamma^\rho \partial_\rho
		\bigg] \gamma_5
		\bigg\}^{\beta\alpha} e^{-i r \cdot x}
		\Bigg|_{r=\omega v}
		\,.
		\label{eq:BLCDAs2pt}
	\end{multline}

	The derivatives $\partial_\mu \equiv \partial/\partial r^\mu$ are understood to act on the hard-scattering kernel.

	The Exponential model for $B$-meson LCDAs is given as 
	\cite{Grozin:1996pq,Braun:2017liq}
		\begin{align}
		\Phi_+(\omega) &= \frac{\omega}{\lambda_{B}^2}e^{-\omega/\lambda_{B}}, \\
		\Phi_-(\omega)
		&= \frac{1}{\lambda_{B}}e^{-\omega/\lambda_{B}} - \frac{\lambda_E^2 - \lambda_H^2}{18 \lambda_{B}^5} \left(2 \lambda_{B}^2 - 4 \omega \lambda_{B} + \omega^2 \right)e^{-\omega/\lambda_{B}}\\
		g_+(\omega)&=
		-\frac{\lambda_E^2}{6\lambda_{B}^2}
		\biggl\{(\omega -2 \lambda_{B})  \text{Ei}\left(-\frac{\omega}{\lambda_{B}}\right) +  
		(\omega +2\lambda_{B}) e^{-\omega/\lambda_{B}}\left(\ln \frac{\omega}{\lambda_{B}}+\gamma_E\right)-2 \omega e^{-\omega/\lambda_{B}}\biggr\}
		\notag\\
		&\quad + \frac{e^{-\omega/\lambda_{B}}}{2{ \lambda_{B}}}\omega^2\biggl\{1 - \frac{1}{36\lambda_{B}^2}(\lambda_E^2- \lambda_H^2)\biggr\}, \label{gplus-model1} \\
		g_-^{WW}(\omega)
		& =  \frac{3 \omega}{4}\,  e^{-\omega/\lambda_{B}} \,.
		\label{eq:gpmWW}
	\end{align}
	Here, $\text{Ei}(x)$ is the exponential integral. The approximate expression for $g_-$ is derived in the Wandzura-Wilczek limit, as mentioned in Ref.~\cite{Gubernari:2018wyi}.
	
	The Local Duality Model for $p = 1$ is given as \cite{Braun:2017liq}
\begin{align}
	\phi_+(\omega,\mu_0)&=
	\frac{3}{4\omega_0^3}\omega\,(2\omega_0-\omega)\,\theta(2\omega_0-\omega)\,,
	\\
	\phi_-(\omega,\mu_0)&=
	\frac{1}{8\omega_0^3}\left[3(2\omega_0-\omega)^2
	-\frac{10(\lambda_E^2-\lambda_H^2)}{3\omega_0^2}\left(3\omega^2-6\omega\omega_0+2\omega_0^2\right)\right]\theta(2\omega_0-\omega),\\
	g_+(\omega,\mu_0)&
	=\frac{5\theta(2\omega_0-\omega)}{384\omega_0^5}\biggl\{\omega(2\omega_0-\omega)
	\Big[8\lambda_E^2(\omega^2-4\omega\omega_0+2\omega_0^2)+\omega(2\omega_0-\omega)(2\lambda_H^2+9\omega_0^2)\Big]
	\notag\\&\quad
	+4\lambda_E^2\biggl[16\omega_0^3(\omega_0-\omega)\,\ln\Big(1-\frac{\omega}{2\omega_0}\Big)
	+\omega^3(4\omega_0-\omega)\,\ln\Big(\frac{2\omega_0}{\omega}-1\Big)
	\biggr]\biggr\},
	\label{gplus-model2a}
\end{align}
	where $\omega_0$ =  $\displaystyle\frac{3}{2}\lambda_B $.

	\section{Coefficients of the LCSR formula} \label{app:BDB}
	Here, we provide a listing of all the coefficients of the two-particle LCDAs that are involved in Eq. \eqref{eq:FF}. \\
	\noindent
	For $f_+^{B \to P}$: 
	
	\begin{equation}
	\begin{aligned}
		C^{(f_+^{B \to P},\phi_+)}_1     & = -\bar{\sigma}\,, &  \\
		C^{(f_+^{B \to P},\bar{\phi})}_2 & = -m_B\bar{\sigma}^2\,, \\
		C^{(f_+^{B \to P},g_+)}_2        & = -4\bar{\sigma}, &
		C^{(f_+^{B \to P},g_+)}_3        & = 8 m_{q_1}^2 \bar{\sigma}\,, \\
		C^{(f_+^{B \to P},\bar{g})}_3    & = -8 m_B \bar{\sigma}^2\,, &
		C^{(f_+^{B \to P},\bar{g})}_4    & = 24 m_{q_1}^2 m_B\bar{\sigma}^2\,.
	\end{aligned}
\end{equation}
For $f_{+/-}^{B \to P}$:
\begin{equation}
	\begin{aligned}
		C^{(f_{+/-}^{B \to P},\phi_+)}_1     & = 2\sigma-1\,,                             & \\
		C^{(f_{+/-}^{B \to P},\bar{\phi})}_2 & = 2 m_B\sigma\bar{\sigma}-m_{q_1}\,,           & \\
		C^{(f_{+/-}^{B \to P},g_+)}_2        & = 4(2\sigma-1)\,,                           &
		C^{(f_{+/-}^{B \to P},g_+)}_3        & = -8 m_{q_1}^2 (2\sigma-1)\,,                     \\
		C^{(f_{+/-}^{B \to P},\bar{g})}_3    & = 16 m_B \sigma\bar{\sigma}\,,             &
		C^{(f_{+/-}^{B \to P},\bar{g})}_4    & = 24 m_{q_1}^2(m_{q_1}-2 m_B \sigma\bar{\sigma})\,.
	\end{aligned}
\end{equation}
	For $f_T^{B \to P}$,

	\begin{equation}
		\begin{aligned}
			&C^{(f_T^{B \to P},\bar{\phi})}_1= \frac{1}{m_B},&
			&C^{(f_T^{B \to P},\bar{\phi})}_2= \frac{-(m_B^2 \bar{\sigma}^2-m_{q_1}^2+2q^2\sigma-q^2)}{m_B},& \\
			&C^{(f_T^{B \to P},\bar{g})}_2= \frac{8}{m_B},&
			&C^{(f_T^{B \to P},\bar{g})}_3= \frac{-8(m_B^2 \bar{\sigma}^2+2 m_{q_1}^2+2q^2\sigma-q^2)}{m_B},&\\
			&C^{(f_T^{B \to P},\bar{g})}_4= \frac{24 m_{q_1}^2(m_B^2 \bar{\sigma}^2-m_{q_1}^2+2q^2\sigma-q^2)}{m_B}.&
		\end{aligned}
	\end{equation}
	
	The normalization factors (in Eq. \eqref{eq:FF}) are,
	\begin{equation}
		K^{(f_+^{B \to P})}=K^{(f_{+/-}^{B \to P})}=f_P, ~~~~~~~
		K^{(f_T^{B \to P})}=\frac{f_P(m^2_B-m^2_P-q^2)}{m_B(m_B+m_P)}\,.
	\end{equation}
	
	\section{$z$-parametrization and fitted coefficients} 
	\label{app:zfit}
	
	In order to extrapolate the form factor in the physical region we first use
	the conformal map from $q^2$- to $z$-plane by
	\begin{equation}
	z(q^2) = \frac{\sqrt{t_+-q^2}-\sqrt{t_+-t_0}}{\sqrt{t_+-q^2}+\sqrt{t_+-t_0}}\,,  
	\end{equation}
	where
	$t_\pm \equiv (m_B\pm m_{P})^2$, with $P = K,\pi,D$ and $t_0\equiv t_+(1-\sqrt{1-t_-/t_+})$. Here, $t_0$ is a free parameter that governs the size of $z$ in the semileptonic phase space. In this work, we have followed the Bharucha-Straub-Zwicky parametrization according to which any form factor can be parametrized as \cite{Bharucha:2015bzk}: 
	\begin{equation}\label{eq:bszexp}
	f_i(q^2) = \frac{1}{1 - q^2/m_{R,i}^2} \sum_k a_k^i \, [z(q^2)-z(0)]^k\,,
	\end{equation}
	where $m_{R,i}$ denotes the mass of sub-threshold resonances compatible with the quantum numbers of the respective form factors and $a_k^i$'s are the coefficients of the expansion. The details of the masses for the respective channels are provided in Table 5 of Ref.~\cite{Gubernari:2018wyi}.
	
	We perform the $\chi^2$-fit of Eq.~\eqref{eq:bszexp} to the LCSR estimates of the form factors at several different $q^2$ points (shown in red and green error bars in Figs.~\ref{fig:B2Kcomp} and \ref{fig:plots}) ranging from large negative to low $q^2$ values. All other inputs are from Table~\ref{tab:Inputs} and Eqs.\eqref{eq:fBval},\,\eqref{eq:lambdaval}, and the $\lambda_B$ value is used either from Eq.~\eqref{eq:lambdaBValue} or Eq.~\eqref{eq:lambdaBValue1} for the corresponding choice of the $B$-meson LCDA model.
	In Table \ref{tab:coefficients}, we present the results for the fitted expansion coefficients, denoted as $a_k^i$, within the two models for the various modes considered in this study. Further details, including the respective correlations among the fitted coefficients obtained in the Exponential (Local Duality) Model, are provided in Tables \ref{tab:corrB2K}, \ref{tab:corrB2D}, and \ref{tab:corrB2Pi} (Tables \ref{tab:corrB2Kdual}, \ref{tab:corrB2Ddual}, and \ref{tab:corrB2Pidual}).
	
	\begin{table*} [h!!!!!]
		\small
		\centering
		\renewcommand*{\arraystretch}{1.3}
		\begin{tabular}{|c||c|c|c|c|c|c|}
			\hline
			& \multicolumn{3}{|c|} {Exponential Model} & \multicolumn{3}{|c|} {Local Duality Model} \\
			\hline\hline
			& $B \to K$ & $B \to \pi$ & $B \to D$ & $B \to K$ & $B \to \pi$ & $B \to D$ \\
			\hline
			$a^+_0$ & 0.39 $\pm$ 0.09 & 0.31 $\pm$ 0.09 & 0.83 $\pm$ 0.11 & 0.38 $\pm$ 0.10 & 0.29 $\pm$ 0.09 & 0.89 $\pm$ 0.09\\
			$a^+_1$ & -0.97 $\pm$ 0.15 & -0.78 $\pm$ 0.17 & -0.50 $\pm$ 0.90 & -1.67 $\pm$ 0.24 & -1.22 $\pm$ 0.28 & -2.58 $\pm$ 1.91\\
			$a^+_2$ & 0.14 $\pm$ 0.53 & -0.09 $\pm$ 0.35 & -2.98 $\pm$ 3.02 & 3.86 $\pm$ 0.91 & 2.22 $\pm$ 0.75 & 1.35 $\pm$ 13.3\\
			$a^0_1$ & 0.47 $\pm$ 0.15 & 0.35 $\pm$ 0.11 & 2.04 $\pm$ 0.69 & -0.01 $\pm$ 0.17 & 0.001 $\pm$ 0.450 & 1.21 $\pm$ 1.15\\
			$a^0_2$ & -0.64 $\pm$ 0.07 & -0.39 $\pm$ 0.06 & -1.80 $\pm$ 0.81 & 0.01 $\pm$ 0.28 & 0.02 $\pm$ 4.48 & -3.35 $\pm$ 7.40\\
			$a^T_0$ & 0.34 $\pm$ 0.06 & 0.27 $\pm$ 0.06 & 0.50 $\pm$ 0.06& 0.38 $\pm$ 0.08 & 0.28 $\pm$ 0.07 & 0.72 $\pm$ 0.16 \\
			$a^T_1$ & -0.90 $\pm$ 0.11 & -0.70 $\pm$ 0.12 & -0.99 $\pm$ 0.72 & -1.79 $\pm$ 0.23 & -1.22 $\pm$ 0.25 & -2.23 $\pm$ 3.08 \\
			$a^T_2$ & 0.37 $\pm$ 0.48 & 0.01 $\pm$ 0.31 & 3.50 $\pm$ 2.41 & 4.43 $\pm$ 1.11 & 2.37 $\pm$ 0.86 & 4.02 $\pm$ 18.7\\
			\hline\hline	\end{tabular}
		\caption{\small $z$-expansion coefficients for the various channels obtained in this work in the two models.}.
		\label{tab:coefficients}
	\end{table*}
	
	\begin{table*} [h!!!!!]
		\small
		\centering
		\renewcommand*{\arraystretch}{1.3}
		\begin{tabular}{|c||cccccccc|}
			\hline
			& $a^+_0$ & $a^+_1$ & $a^+_2$ & $a^0_1$ & $a^0_2$ & $a^T_0$ & $a^T_1$ & $a^T_2$ \\
			\hline\hline
			$a^+_0$ & $1.$  &  $-0.743$  &  $-0.677$  &  $0.867$  &  $-0.297$  &  $0.984$  &  $-0.327$  &  $-0.689$  \\
			$a^+_1$ & $-0.743$  &  $1.$  &  $0.014$  &  $-0.350$  &  $-0.054$  &  $-0.815$  &  $0.862$  &  $0.040$  \\
			$a^+_2$ & $-0.677$  &  $0.014$  &  $1.$  &  $-0.904$  &  $0.549$  &  $-0.573$  &  $-0.460$  &  $0.995$  \\
			$a^0_1$ & $0.867$  &  $-0.350$  &  $-0.904$  &  $1.$  &  $-0.505$  &  $0.827$  &  $0.084$  &  $-0.898$  \\
			$a^0_2$ & $-0.297$  &  $-0.054$  &  $0.549$  &  $-0.505$  &  $1.$  &  $-0.274$  &  $-0.223$  &  $0.524$  \\
			$a^T_0$ & $0.984$  &  $-0.815$  &  $-0.573$  &  $0.827$  &  $-0.274$  &  $1.$  &  $-0.459$  &  $-0.586$  \\
			$a^T_1$ & $-0.327$  &  $0.862$  &  $-0.460$  &  $0.084$  &  $-0.223$  &  $-0.459$  &  $1.$  &  $-0.445$  \\
			$a^T_2$ & $-0.689$  &  $0.040$  &  $0.995$  &  $-0.898$  &  $0.524$  &  $-0.586$  &  $-0.445$  &  $1.$  \\
			\hline\hline
		\end{tabular}
		\caption{\small Correlation among the $z$-expansion coefficients for the $B \to K$ mode in the Exponential Model.} 
		\label{tab:corrB2K}
	\end{table*}
	
	\begin{table*} [h!!!!!]
		\small
		\centering
		\renewcommand*{\arraystretch}{1.3}
		\begin{tabular}{|c||cccccccc|}
			\hline
			& $a^+_0$ & $a^+_1$ & $a^+_2$ & $a^0_1$ & $a^0_2$ & $a^T_0$ & $a^T_1$ & $a^T_2$ \\
			\hline\hline
			$a^+_0$ & $1.$  &  $0.849$  &  $-0.902$  &  $0.928$  &  $-0.194$  &  $-0.498$  &  $0.875$  &  $-0.823$  \\
			$a^+_1$ & $0.849$  &  $1.$  &  $-0.889$  &  $0.98$  &  $0.027$  &  $-0.853$  &  $0.920$  &  $-0.700$  \\
			$a^+_2$ & $-0.902$  &  $-0.889$  &  $1.$  &  $-0.919$  &  $0.393$  &  $0.696$  &  $-0.976$  &  $0.907$  \\
			$a^0_1$  & $0.928$  &  $0.98$  &  $-0.919$  &  $1.$  &  $-0.036$  &  $-0.742$  &  $0.923$  &  $-0.741$  \\
			$a^0_2$ & $-0.194$  &  $0.027$  &  $0.393$  &  $-0.036$  &  $1.$  &  $-0.048$  &  $-0.272$  &  $0.533$  \\
			$a^T_0$ & $-0.498$  &  $-0.853$  &  $0.696$  &  $-0.742$  &  $-0.048$  &  $1.$  &  $-0.788$  &  $0.535$  \\
			$a^T_1$ & $0.875$  &  $0.920$  &  $-0.976$  &  $0.923$  &  $-0.272$  &  $-0.788$  &  $1.$  &  $-0.907$  \\
			$a^T_2$ & $-0.823$  &  $-0.700$  &  $0.907$  &  $-0.741$  &  $0.533$  &  $0.535$  &  $-0.907$  &  $1.$  \\
			\hline\hline
		\end{tabular}
		\caption{\small Correlation among the $z$-expansion coefficients for the $B \to D$ mode in the Exponential Model.} 
		\label{tab:corrB2D}
	\end{table*}
	
	\begin{table*} [h!!!!!]
		\small
		\centering
		\renewcommand*{\arraystretch}{1.3}
		\begin{tabular}{|c||cccccccc|}
			\hline
			& $a^+_0$ & $a^+_1$ & $a^+_2$ & $a^0_1$ & $a^0_2$ & $a^T_0$ & $a^T_1$ & $a^T_2$ \\
			\hline\hline
			$a^+_0$ & $1.$  &  $-0.897$  &  $-0.632$  &  $0.878$  &  $-0.189$  &  $0.990$  &  $-0.744$  &  $-0.636$  \\
			$a^+_1$ & $-0.897$  &  $1.$  &  $0.227$  &  $-0.610$  &  $0.111$  &  $-0.932$  &  $0.956$  &  $0.237$  \\
			$a^+_2$ & $-0.632$  &  $0.227$  &  $1.$  &  $-0.863$  &  $0.297$  &  $-0.549$  &  $-0.032$  &  $0.994$  \\
			$a^0_1$ & $0.878$  &  $-0.610$  &  $-0.863$  &  $1.$  &  $-0.284$  &  $0.856$  &  $-0.424$  &  $-0.859$  \\
			$a^0_2$ & $-0.189$  &  $0.111$  &  $0.297$  &  $-0.284$  &  $1.$  &  $-0.197$  &  $0.114$  &  $0.257$  \\
			$a^T_0$ & $0.990$  &  $-0.932$  &  $-0.549$  &  $0.856$  &  $-0.197$  &  $1.$  &  $-0.814$  &  $-0.554$  \\
			$a^T_1$ & $-0.744$  &  $0.956$  &  $-0.032$  &  $-0.424$  &  $0.114$  &  $-0.814$  &  $1.$  &  $-0.027$  \\
			$a^T_2$ & $-0.636$  &  $0.237$  &  $0.994$  &  $-0.859$  &  $0.257$  &  $-0.554$  &  $-0.027$  &  $1.$  \\
			\hline\hline
		\end{tabular}
		\caption{\small Correlation among the $z$-expansion coefficients for the $B \to \pi$ mode in the Exponential Model.} 
		\label{tab:corrB2Pi}
	\end{table*}
	
	\begin{table*} [h!!!!!]
		\small
		\centering
		\renewcommand*{\arraystretch}{1.3}
		\begin{tabular}{|c||cccccccc|}
			\hline
			& $a^+_0$ & $a^+_1$ & $a^+_2$ & $a^0_1$ & $a^0_2$ & $a^T_0$ & $a^T_1$ & $a^T_2$ \\
			\hline\hline
			$a^+_0$ & $1.$  &  $-0.757$  &  $-0.473$  &  $0.78$  &  $-0.884$  &  $0.985$  &  $-0.405$  &  $-0.589$  \\
			$a^+_1$ & $-0.757$  &  $1.$  &  $-0.213$  &  $-0.208$  &  $0.442$  &  $-0.834$  &  $0.895$  &  $-0.067$  \\
			$a^+_2$ & $-0.473$  &  $-0.213$  &  $1.$  &  $-0.871$  &  $0.736$  &  $-0.346$  &  $-0.604$  &  $0.985$  \\
			$a^0_1$ & $0.78$  &  $-0.208$  &  $-0.871$  &  $1.$  &  $-0.934$  &  $0.712$  &  $0.176$  &  $-0.897$  \\
			$a^0_2$ & $-0.884$  &  $0.442$  &  $0.736$  &  $-0.934$  &  $1.$  &  $-0.846$  &  $0.07$  &  $0.805$  \\
			$a^T_0$ & $0.985$  &  $-0.834$  &  $-0.346$  &  $0.712$  &  $-0.846$  &  $1.$  &  $-0.536$  &  $-0.468$  \\
			$a^T_1$ & $-0.405$  &  $0.895$  &  $-0.604$  &  $0.176$  &  $0.07$  &  $-0.536$  &  $1.$  &  $-0.489$  \\
			$a^T_2$ & $-0.589$  &  $-0.067$  &  $0.985$  &  $-0.897$  &  $0.805$  &  $-0.468$  &  $-0.489$  &  $1.$  \\
			\hline\hline
		\end{tabular}
		\caption{\small Correlation among the $z$-expansion coefficients for the $B \to K$ mode in the Local Duality Model.}
		\label{tab:corrB2Kdual}
	\end{table*}
	
	\begin{table*} [h!!!!!]
		\small
		\centering
		\renewcommand*{\arraystretch}{1.3}
		\begin{tabular}{|c||cccccccc|}
			\hline
			& $a^+_0$ & $a^+_1$ & $a^+_2$ & $a^0_1$ & $a^0_2$ & $a^T_0$ & $a^T_1$ & $a^T_2$ \\
			\hline\hline
			$a^+_0$ & 
			$1.$  &  $0.875$  &  $-0.881$  &  $0.713$  &  $-0.051$  &  $-0.797$  &  $0.802$  &  $-0.549$  \\
			$a^+_1$ & $0.875$  &  $1.$  &  $-0.971$  &  $0.844$  &  $-0.155$  &  $-0.972$  &  $0.832$  &  $-0.477$  \\
			$a^+_2$ & $-0.881$  &  $-0.971$  &  $1.$  &  $-0.919$  &  $0.344$  &  $0.969$  &  $-0.921$  &  $0.632$  \\
			$a^0_1$  &  $0.713$  &  $0.844$  &  $-0.919$  &  $1.$  &  $-0.642$  &  $-0.867$  &  $0.848$  &  $-0.581$  \\
			$a^0_2$ &  $-0.051$  &  $-0.155$  &  $0.344$  &  $-0.642$  &  $1.$  &  $0.262$  &  $-0.419$  &  $0.457$  \\
			$a^T_0$  & $-0.797$  &  $-0.972$  &  $0.969$  &  $-0.867$  &  $0.262$  &  $1.$  &  $-0.884$  &  $0.568$  \\
			$a^T_1$ & $0.802$  &  $0.832$  &  $-0.921$  &  $0.848$  &  $-0.419$  &  $-0.884$  &  $1.$  &  $-0.876$  \\
			$a^T_2$ & $-0.549$  &  $-0.477$  &  $0.632$  &  $-0.581$  &  $0.457$  &  $0.568$  &  $-0.876$  &  $1.$  \\
			\hline\hline
		\end{tabular}
		\caption{\small Correlation among the $z$-expansion coefficients for the $B \to D$ mode in the Local Duality Model.}
		\label{tab:corrB2Ddual}
	\end{table*}
	
	\begin{table*} [h!!!!!]
		\small
		\centering
		\renewcommand*{\arraystretch}{1.3}
		\begin{tabular}{|c||cccccccc|}
			\hline
			& $a^+_0$ & $a^+_1$ & $a^+_2$ & $a^0_1$ & $a^0_2$ & $a^T_0$ & $a^T_1$ & $a^T_2$ \\
			\hline\hline
			$a^+_0$ & 
			$1.$  &  $-0.621$  &  $-0.226$  &  $0.3$  &  $-0.032$  &  $0.975$  &  $-0.509$  &  $-0.224$  \\
			$a^+_1$  & $-0.621$  &  $1.$  &  $-0.312$  &  $0.244$  &  $-0.118$  &  $-0.748$  &  $0.967$  &  $-0.194$  \\
			$a^+_2$ & $-0.226$  &  $-0.312$  &  $1.$  &  $-0.192$  &  $-0.101$  &  $-0.176$  &  $-0.507$  &  $0.821$  \\
			$a^0_1$  & $0.3$  &  $0.244$  &  $-0.192$  &  $1.$  &  $-0.868$  &  $0.198$  &  $0.279$  &  $-0.52$  \\
			$a^0_2$ & $-0.032$  &  $-0.118$  &  $-0.101$  &  $-0.868$  &  $1.$  &  $0.001$  &  $-0.107$  &  $0.388$  \\
			$a^T_0$  &  $0.975$  &  $-0.748$  &  $-0.176$  &  $0.198$  &  $0.001$  &  $1.$  &  $-0.635$  &  $-0.209$  \\
			$a^T_1$ &  $-0.509$  &  $0.967$  &  $-0.507$  &  $0.279$  &  $-0.107$  &  $-0.635$  &  $1.$  &  $-0.4$  \\
			$a^T_2$ &  $-0.224$  &  $-0.194$  &  $0.821$  &  $-0.52$  &  $0.388$  &  $-0.209$  &  $-0.4$  &  $1.$  \\
			\hline\hline
		\end{tabular}
		\caption{\small Correlation among the $z$-expansion coefficients for the $B \to \pi$ mode in the Local Duality Model.}
		\label{tab:corrB2Pidual}
	\end{table*}


\bibliographystyle{bibstyle} 
%\biboptions{sort&compress}
\bibliography{example.bib}

\begin{thebibliography}{10}
\expandafter\ifx\csname url\endcsname\relax
  \def\url#1{\texttt{#1}}\fi
\expandafter\ifx\csname urlprefix\endcsname\relax\def\urlprefix{URL }\fi
\expandafter\ifx\csname href\endcsname\relax
  \def\href#1#2{#2} \def\path#1{#1}\fi

\bibitem{Boyd:1994tt}
C.~G. Boyd, B.~Grinstein, R.~F. Lebed, {Constraints on form-factors for
  exclusive semileptonic heavy to light meson decays}, Phys. Rev. Lett. 74
  (1995) 4603--4606.
\newblock \href {http://arxiv.org/abs/hep-ph/9412324}
  {\path{arXiv:hep-ph/9412324}}, \href
  {https://doi.org/10.1103/PhysRevLett.74.4603}
  {\path{doi:10.1103/PhysRevLett.74.4603}}.

\bibitem{DiCarlo:2021dzg}
M.~Di~Carlo, G.~Martinelli, M.~Naviglio, F.~Sanfilippo, S.~Simula, L.~Vittorio,
  {Unitarity bounds for semileptonic decays in lattice QCD}, Phys. Rev. D
  104~(5) (2021) 054502.
\newblock \href {http://arxiv.org/abs/2105.02497} {\path{arXiv:2105.02497}},
  \href {https://doi.org/10.1103/PhysRevD.104.054502}
  {\path{doi:10.1103/PhysRevD.104.054502}}.

\bibitem{Flynn:2023qmi}
J.~M. Flynn, A.~J\"uttner, J.~T. Tsang, {Bayesian inference for form-factor
  fits regulated by unitarity and analyticity} (3 2023).
\newblock \href {http://arxiv.org/abs/2303.11285} {\path{arXiv:2303.11285}}.

\bibitem{Braun:2003wx}
V.~M. Braun, D.~Y. Ivanov, G.~P. Korchemsky, {The B meson distribution
  amplitude in QCD}, Phys. Rev. D 69 (2004) 034014.
\newblock \href {http://arxiv.org/abs/hep-ph/0309330}
  {\path{arXiv:hep-ph/0309330}}, \href
  {https://doi.org/10.1103/PhysRevD.69.034014}
  {\path{doi:10.1103/PhysRevD.69.034014}}.

\bibitem{Khodjamirian:2020hob}
A.~Khodjamirian, R.~Mandal, T.~Mannel, {Inverse moment of the B$_{s}$-meson
  distribution amplitude from QCD sum rule}, JHEP 10 (2020) 043.
\newblock \href {http://arxiv.org/abs/2008.03935} {\path{arXiv:2008.03935}},
  \href {https://doi.org/10.1007/JHEP10(2020)043}
  {\path{doi:10.1007/JHEP10(2020)043}}.

\bibitem{Belle:2018jqd}
M.~Gelb, et~al., {Search for the rare decay of $B^+ \to \ell^{\,+} \nu_{\ell}
  \gamma$ with improved hadronic tagging}, Phys. Rev. D 98~(11) (2018) 112016.
\newblock \href {http://arxiv.org/abs/1810.12976} {\path{arXiv:1810.12976}},
  \href {https://doi.org/10.1103/PhysRevD.98.112016}
  {\path{doi:10.1103/PhysRevD.98.112016}}.

\bibitem{Giusti:2023pot}
D.~Giusti, C.~F. Kane, C.~Lehner, S.~Meinel, A.~Soni, {Methods for
  high-precision determinations of radiative-leptonic decay form factors using
  lattice QCD}, Phys. Rev. D 107~(7) (2023) 074507.
\newblock \href {http://arxiv.org/abs/2302.01298} {\path{arXiv:2302.01298}},
  \href {https://doi.org/10.1103/PhysRevD.107.074507}
  {\path{doi:10.1103/PhysRevD.107.074507}}.

\bibitem{Desiderio:2020oej}
A.~Desiderio, et~al., {First lattice calculation of radiative leptonic decay
  rates of pseudoscalar mesons}, Phys. Rev. D 103~(1) (2021) 014502.
\newblock \href {http://arxiv.org/abs/2006.05358} {\path{arXiv:2006.05358}},
  \href {https://doi.org/10.1103/PhysRevD.103.014502}
  {\path{doi:10.1103/PhysRevD.103.014502}}.

\bibitem{Frezzotti:2023ygt}
R.~Frezzotti, N.~Tantalo, G.~Gagliardi, F.~Sanfilippo, S.~Simula, V.~Lubicz,
  F.~Mazzetti, G.~Martinelli, C.~T. Sachrajda, {Lattice calculation of the Ds
  meson radiative form factors over the full kinematical range}, Phys. Rev. D
  108~(7) (2023) 074505.
\newblock \href {http://arxiv.org/abs/2306.05904} {\path{arXiv:2306.05904}},
  \href {https://doi.org/10.1103/PhysRevD.108.074505}
  {\path{doi:10.1103/PhysRevD.108.074505}}.

\bibitem{Wang:2015vgv}
Y.-M. Wang, Y.-L. Shen, {QCD corrections to $B \to pi $ form factors from
  light-cone sum rules}, Nucl. Phys. B 898 (2015) 563--604.
\newblock \href {http://arxiv.org/abs/1506.00667} {\path{arXiv:1506.00667}},
  \href {https://doi.org/10.1016/j.nuclphysb.2015.07.016}
  {\path{doi:10.1016/j.nuclphysb.2015.07.016}}.

\bibitem{Janowski:2021yvz}
T.~Janowski, B.~Pullin, R.~Zwicky, {Charged and neutral $
  {\overline{B}}_{u,d,s} $ $\to \gamma$ form factors from light cone sum rules
  at NLO}, JHEP 12 (2021) 008.
\newblock \href {http://arxiv.org/abs/2106.13616} {\path{arXiv:2106.13616}},
  \href {https://doi.org/10.1007/JHEP12(2021)008}
  {\path{doi:10.1007/JHEP12(2021)008}}.

\bibitem{Gao:2019lta}
J.~Gao, C.-D. L\"u, Y.-L. Shen, Y.-M. Wang, Y.-B. Wei, {Precision calculations
  of $B \to V$ form factors from soft-collinear effective theory sum rules on
  the light-cone}, Phys. Rev. D 101~(7) (2020) 074035.
\newblock \href {http://arxiv.org/abs/1907.11092} {\path{arXiv:1907.11092}},
  \href {https://doi.org/10.1103/PhysRevD.101.074035}
  {\path{doi:10.1103/PhysRevD.101.074035}}.

\bibitem{Parrott:2022rgu}
W.~G. Parrott, C.~Bouchard, C.~T.~H. Davies, {$B \to K$ and $ D \to K$ form
  factors from fully relativistic lattice QCD}, Phys. Rev. D 107~(1) (2023)
  014510.
\newblock \href {http://arxiv.org/abs/2207.12468} {\path{arXiv:2207.12468}},
  \href {https://doi.org/10.1103/PhysRevD.107.014510}
  {\path{doi:10.1103/PhysRevD.107.014510}}.

\bibitem{Bailey:2015dka}
J.~A. Bailey, et~al., {$B\to Kl^+l^-$ Decay Form Factors from Three-Flavor
  Lattice QCD}, Phys. Rev. D 93~(2) (2016) 025026.
\newblock \href {http://arxiv.org/abs/1509.06235} {\path{arXiv:1509.06235}},
  \href {https://doi.org/10.1103/PhysRevD.93.025026}
  {\path{doi:10.1103/PhysRevD.93.025026}}.

\bibitem{Bouchard:2013eph}
C.~Bouchard, G.~P. Lepage, C.~Monahan, H.~Na, J.~Shigemitsu, {Rare decay $B \to
  K \ell^+ \ell^-$ form factors from lattice QCD}, Phys. Rev. D 88~(5) (2013)
  054509, [Erratum: Phys.Rev.D 88, 079901 (2013)].
\newblock \href {http://arxiv.org/abs/1306.2384} {\path{arXiv:1306.2384}},
  \href {https://doi.org/10.1103/PhysRevD.88.054509}
  {\path{doi:10.1103/PhysRevD.88.054509}}.

\bibitem{Faller:2008tr}
S.~Faller, A.~Khodjamirian, C.~Klein, T.~Mannel, { $B \to D^{(*)}$ Form Factors
  from QCD Light-Cone Sum Rules}, Eur. Phys. J. C 60 (2009) 603--615.
\newblock \href {http://arxiv.org/abs/0809.0222} {\path{arXiv:0809.0222}},
  \href {https://doi.org/10.1140/epjc/s10052-009-0968-4}
  {\path{doi:10.1140/epjc/s10052-009-0968-4}}.

\bibitem{FlavourLatticeAveragingGroupFLAG:2021npn}
Y.~Aoki, et~al., {FLAG Review 2021}, Eur. Phys. J. C 82~(10) (2022) 869.
\newblock \href {http://arxiv.org/abs/2111.09849} {\path{arXiv:2111.09849}},
  \href {https://doi.org/10.1140/epjc/s10052-022-10536-1}
  {\path{doi:10.1140/epjc/s10052-022-10536-1}}.

\bibitem{RBC:2014ntl}
T.~Blum, et~al., {Domain wall QCD with physical quark masses}, Phys. Rev. D
  93~(7) (2016) 074505.
\newblock \href {http://arxiv.org/abs/1411.7017} {\path{arXiv:1411.7017}},
  \href {https://doi.org/10.1103/PhysRevD.93.074505}
  {\path{doi:10.1103/PhysRevD.93.074505}}.

\bibitem{Follana:2007uv}
E.~Follana, C.~T.~H. Davies, G.~P. Lepage, J.~Shigemitsu, {High Precision
  determination of the pi, K, D and D(s) decay constants from lattice QCD},
  Phys. Rev. Lett. 100 (2008) 062002.
\newblock \href {http://arxiv.org/abs/0706.1726} {\path{arXiv:0706.1726}},
  \href {https://doi.org/10.1103/PhysRevLett.100.062002}
  {\path{doi:10.1103/PhysRevLett.100.062002}}.

\bibitem{MILC:2010hzw}
A.~Bazavov, et~al., {Results for light pseudoscalar mesons}, PoS LATTICE2010
  (2010) 074.
\newblock \href {http://arxiv.org/abs/1012.0868} {\path{arXiv:1012.0868}},
  \href {https://doi.org/10.22323/1.105.0074} {\path{doi:10.22323/1.105.0074}}.

\bibitem{Khodjamirian:2003xk}
A.~Khodjamirian, T.~Mannel, M.~Melcher, {Flavor SU(3) symmetry in charmless B
  decays}, Phys. Rev. D 68 (2003) 114007.
\newblock \href {http://arxiv.org/abs/hep-ph/0308297}
  {\path{arXiv:hep-ph/0308297}}, \href
  {https://doi.org/10.1103/PhysRevD.68.114007}
  {\path{doi:10.1103/PhysRevD.68.114007}}.

\bibitem{Colangelo:2000dp}
P.~Colangelo, A.~Khodjamirian, {QCD sum rules, a modern perspective} (2000)
  1495--1576\href {http://arxiv.org/abs/hep-ph/0010175}
  {\path{arXiv:hep-ph/0010175}}, \href
  {https://doi.org/10.1142/9789812810458_0033}
  {\path{doi:10.1142/9789812810458_0033}}.

\bibitem{Khodjamirian:2006st}
A.~Khodjamirian, T.~Mannel, N.~Offen, {Form-factors from light-cone sum rules
  with B-meson distribution amplitudes}, Phys. Rev. D 75 (2007) 054013.
\newblock \href {http://arxiv.org/abs/hep-ph/0611193}
  {\path{arXiv:hep-ph/0611193}}, \href
  {https://doi.org/10.1103/PhysRevD.75.054013}
  {\path{doi:10.1103/PhysRevD.75.054013}}.

\bibitem{FermilabLattice:2014tsy}
A.~Bazavov, et~al., {Charmed and Light Pseudoscalar Meson Decay Constants from
  Four-Flavor Lattice QCD with Physical Light Quarks}, Phys. Rev. D 90~(7)
  (2014) 074509.
\newblock \href {http://arxiv.org/abs/1407.3772} {\path{arXiv:1407.3772}},
  \href {https://doi.org/10.1103/PhysRevD.90.074509}
  {\path{doi:10.1103/PhysRevD.90.074509}}.

\bibitem{Dowdall:2013rya}
R.~J. Dowdall, C.~T.~H. Davies, G.~P. Lepage, C.~McNeile, {Vus from pi and K
  decay constants in full lattice QCD with physical u, d, s and c quarks},
  Phys. Rev. D 88 (2013) 074504.
\newblock \href {http://arxiv.org/abs/1303.1670} {\path{arXiv:1303.1670}},
  \href {https://doi.org/10.1103/PhysRevD.88.074504}
  {\path{doi:10.1103/PhysRevD.88.074504}}.

\bibitem{Carrasco:2014poa}
N.~Carrasco, et~al., {Leptonic decay constants $f_{K},f_{D},$ and $f_{{D}_{s}}$
  with $N_{f} = 2+1+1$ twisted-mass lattice QCD}, Phys. Rev. D 91~(5) (2015)
  054507.
\newblock \href {http://arxiv.org/abs/1411.7908} {\path{arXiv:1411.7908}},
  \href {https://doi.org/10.1103/PhysRevD.91.054507}
  {\path{doi:10.1103/PhysRevD.91.054507}}.

\bibitem{Bazavov:2017lyh}
A.~Bazavov, et~al., {$B$- and $D$-meson leptonic decay constants from
  four-flavor lattice QCD}, Phys. Rev. D 98~(7) (2018) 074512.
\newblock \href {http://arxiv.org/abs/1712.09262} {\path{arXiv:1712.09262}},
  \href {https://doi.org/10.1103/PhysRevD.98.074512}
  {\path{doi:10.1103/PhysRevD.98.074512}}.

\bibitem{Gubernari:2018wyi}
N.~Gubernari, A.~Kokulu, D.~van Dyk, {$B\to P$ and $B\to V$ Form Factors from
  $B$-Meson Light-Cone Sum Rules beyond Leading Twist}, JHEP 01 (2019) 150.
\newblock \href {http://arxiv.org/abs/1811.00983} {\path{arXiv:1811.00983}},
  \href {https://doi.org/10.1007/JHEP01(2019)150}
  {\path{doi:10.1007/JHEP01(2019)150}}.

\bibitem{Gubernari:2022hrq}
N.~Gubernari, A.~Khodjamirian, R.~Mandal, T.~Mannel, {$B \to D_{1}(2420)$ and
  $B \to {\mathrm{D}}_1^{\prime } (2430)$ form factors from QCD light-cone sum
  rules}, JHEP 05 (2022) 029.
\newblock \href {http://arxiv.org/abs/2203.08493} {\path{arXiv:2203.08493}},
  \href {https://doi.org/10.1007/JHEP05(2022)029}
  {\path{doi:10.1007/JHEP05(2022)029}}.

\bibitem{Braun:2017liq}
V.~M. Braun, Y.~Ji, A.~N. Manashov, {Higher-twist B-meson Distribution
  Amplitudes in HQET}, JHEP 05 (2017) 022.
\newblock \href {http://arxiv.org/abs/1703.02446} {\path{arXiv:1703.02446}},
  \href {https://doi.org/10.1007/JHEP05(2017)022}
  {\path{doi:10.1007/JHEP05(2017)022}}.

\bibitem{Nishikawa:2011qk}
T.~Nishikawa, K.~Tanaka, {QCD Sum Rules for Quark-Gluon Three-Body Components
  in the B Meson}, Nucl. Phys. B 879 (2014) 110--142.
\newblock \href {http://arxiv.org/abs/1109.6786} {\path{arXiv:1109.6786}},
  \href {https://doi.org/10.1016/j.nuclphysb.2013.12.007}
  {\path{doi:10.1016/j.nuclphysb.2013.12.007}}.

\bibitem{ETM:2016nbo}
A.~Bussone, et~al., {Mass of the b quark and B -meson decay constants from
  N$_f$=2+1+1 twisted-mass lattice QCD}, Phys. Rev. D 93~(11) (2016) 114505.
\newblock \href {http://arxiv.org/abs/1603.04306} {\path{arXiv:1603.04306}},
  \href {https://doi.org/10.1103/PhysRevD.93.114505}
  {\path{doi:10.1103/PhysRevD.93.114505}}.

\bibitem{Dowdall:2013tga}
R.~J. Dowdall, C.~T.~H. Davies, R.~R. Horgan, C.~J. Monahan, J.~Shigemitsu,
  {B-Meson Decay Constants from Improved Lattice Nonrelativistic QCD with
  Physical u, d, s, and c Quarks}, Phys. Rev. Lett. 110~(22) (2013) 222003.
\newblock \href {http://arxiv.org/abs/1302.2644} {\path{arXiv:1302.2644}},
  \href {https://doi.org/10.1103/PhysRevLett.110.222003}
  {\path{doi:10.1103/PhysRevLett.110.222003}}.

\bibitem{Hughes:2017spc}
C.~Hughes, C.~T.~H. Davies, C.~J. Monahan, {New methods for B meson decay
  constants and form factors from lattice NRQCD}, Phys. Rev. D 97~(5) (2018)
  054509.
\newblock \href {http://arxiv.org/abs/1711.09981} {\path{arXiv:1711.09981}},
  \href {https://doi.org/10.1103/PhysRevD.97.054509}
  {\path{doi:10.1103/PhysRevD.97.054509}}.

\bibitem{Ball:2004ye}
P.~Ball, R.~Zwicky, {New results on $B \to \pi, K, \eta$ decay formfactors from
  light-cone sum rules}, Phys. Rev. D 71 (2005) 014015.
\newblock \href {http://arxiv.org/abs/hep-ph/0406232}
  {\path{arXiv:hep-ph/0406232}}, \href
  {https://doi.org/10.1103/PhysRevD.71.014015}
  {\path{doi:10.1103/PhysRevD.71.014015}}.

\bibitem{Lange:2003ff}
B.~O. Lange, M.~Neubert, {Renormalization group evolution of the B meson light
  cone distribution amplitude}, Phys. Rev. Lett. 91 (2003) 102001.
\newblock \href {http://arxiv.org/abs/hep-ph/0303082}
  {\path{arXiv:hep-ph/0303082}}, \href
  {https://doi.org/10.1103/PhysRevLett.91.102001}
  {\path{doi:10.1103/PhysRevLett.91.102001}}.

\bibitem{MILC:2015uhg}
J.~A. Bailey, et~al., {B\textrightarrow{}D\ensuremath{\ell}\ensuremath{\nu}
  form factors at nonzero recoil and |V$_{cb}$| from 2+1-flavor lattice QCD},
  Phys. Rev. D 92~(3) (2015) 034506.
\newblock \href {http://arxiv.org/abs/1503.07237} {\path{arXiv:1503.07237}},
  \href {https://doi.org/10.1103/PhysRevD.92.034506}
  {\path{doi:10.1103/PhysRevD.92.034506}}.

\bibitem{FermilabLattice:2015mwy}
J.~A. Bailey, et~al., {$|V_{ub}|$ from $B\to\pi\ell\nu$ decays and (2+1)-flavor
  lattice QCD}, Phys. Rev. D 92~(1) (2015) 014024.
\newblock \href {http://arxiv.org/abs/1503.07839} {\path{arXiv:1503.07839}},
  \href {https://doi.org/10.1103/PhysRevD.92.014024}
  {\path{doi:10.1103/PhysRevD.92.014024}}.

\bibitem{FermilabLattice:2015cdh}
J.~A. Bailey, et~al., {$B\to\pi\ell\ell$ form factors for new-physics searches
  from lattice QCD}, Phys. Rev. Lett. 115~(15) (2015) 152002.
\newblock \href {http://arxiv.org/abs/1507.01618} {\path{arXiv:1507.01618}},
  \href {https://doi.org/10.1103/PhysRevLett.115.152002}
  {\path{doi:10.1103/PhysRevLett.115.152002}}.

\bibitem{Leljak:2021vte}
D.~Leljak, B.~Meli\'c, D.~van Dyk, {The $ \bar{B} \to \pi$ form factors from
  QCD and their impact on $|V_{ub}|$}, JHEP 07 (2021) 036.
\newblock \href {http://arxiv.org/abs/2102.07233} {\path{arXiv:2102.07233}},
  \href {https://doi.org/10.1007/JHEP07(2021)036}
  {\path{doi:10.1007/JHEP07(2021)036}}.

\bibitem{Beneke:2018wjp}
M.~Beneke, V.~M. Braun, Y.~Ji, Y.-B. Wei, {Radiative leptonic decay $B\to
  \gamma \ell \nu_\ell$ with subleading power corrections}, JHEP 07 (2018) 154.
\newblock \href {http://arxiv.org/abs/1804.04962} {\path{arXiv:1804.04962}},
  \href {https://doi.org/10.1007/JHEP07(2018)154}
  {\path{doi:10.1007/JHEP07(2018)154}}.

\bibitem{Grozin:1996pq}
A.~G. Grozin, M.~Neubert, {Asymptotics of heavy meson form-factors}, Phys. Rev.
  D 55 (1997) 272--290.
\newblock \href {http://arxiv.org/abs/hep-ph/9607366}
  {\path{arXiv:hep-ph/9607366}}, \href
  {https://doi.org/10.1103/PhysRevD.55.272}
  {\path{doi:10.1103/PhysRevD.55.272}}.

\bibitem{Bharucha:2015bzk}
A.~Bharucha, D.~M. Straub, R.~Zwicky, {$B\to V\ell^+\ell^-$ in the Standard
  Model from light-cone sum rules}, JHEP 08 (2016) 098.
\newblock \href {http://arxiv.org/abs/1503.05534} {\path{arXiv:1503.05534}},
  \href {https://doi.org/10.1007/JHEP08(2016)098}
  {\path{doi:10.1007/JHEP08(2016)098}}.

\end{thebibliography}



%% For example:

%% \bibitem[Aladro et al.(2015)]{Aladro15} Aladro, R., Martín, S., Riquelme, D., et al. 2015, \aas, 579, A101

%%\end{thebibliography}

\end{document}